\documentclass[preprint,12pt]{elsarticle}

\usepackage{amssymb}

\journal{Nuclear Instruments and Methods in Physics Research A}

\begin{document}

\begin{frontmatter}



\title{The Prototype GAPS (pGAPS) Experiment}


\author[ucla]{S.~A.~I.~Mognet\corref{cor1}}
\ead{mognet@astro.ucla.edu}
\author[columbia]{T.~Aramaki}
\author[isas]{N.~Bando}
\author[cal]{S.~E.~Boggs}
\author[cal]{P.~von~Doetinchem}
\author[isas]{H.~Fuke}
\author[columbia]{F.~H.~Gahbauer}
\author[columbia]{C.~J.~Hailey}
\author[columbia]{J.~E.~Koglin}
\author[columbia]{N.~Madden}
\author[isas]{K.~Mori}
\author[isas]{S.~Okazaki}
\author[ucla]{R.~A.~Ong}
\author[columbia]{K.~M.~Perez}
\author[columbia]{G.~Tajiri}
\author[isas]{T.~Yoshida}
\author[ucla]{J.~Zweerink}

\cortext[cor1]{Corresponding author: Phone: 310-794-9455}

\address[ucla]{Department of Physics and Astronomy, University of California, Los Angeles, CA 90095, USA}
\address[columbia]{Department of Physics, Columbia University, New York, NY 10027, USA}
\address[isas]{Institute of Space and Astronautical Science, Japan Aerospace Exploration Agency
(ISAS/JAXA), Sagamihara, Kanagawa 252-5210, Japan}
\address[cal]{Space Sciences Laboratory, University of California, Berkeley, CA 94720, USA}

\begin{abstract}
The General Antiparticle Spectrometer (GAPS) experiment is a novel approach for the detection of cosmic ray antiparticles. A prototype GAPS experiment (pGAPS) was successfully flown on a high-altitude balloon in June of 2012.  The goals of the pGAPS experiment were: to test the operation of lithium drifted silicon (Si(Li)) detectors at balloon altitudes, to validate the thermal model and cooling concept needed for engineering of a full-size GAPS instrument, and to characterize cosmic ray and X-ray backgrounds. The instrument was launched from the Japan Aerospace Exploration Agency's (JAXA) Taiki Aerospace Research Field in Hokkaido, Japan. The flight lasted a total of 6 hours, with over 3 hours at float altitude (${\sim}33$ km). Over one million cosmic ray triggers were recorded and all flight goals were met or exceeded.
\end{abstract}

\begin{keyword}
Dark Matter \sep Antideuteron \sep Indirect Detection \sep pGAPS \sep bGAPS \sep GAPS

\end{keyword}

\end{frontmatter}


\section{Introduction}
One of the great unanswered questions in physics and astronomy concerns the properties and composition of dark matter. Multiple (and independent) lines of evidence strongly suggest that there exist large quantities of mass in our universe beyond normal, baryonic matter. Currently, dark matter is most successfully explained by postulating some as-yet undiscovered particle (or particles). Numerous theories produce candidate dark matter particles. Some theories postulate very light particles, such as axions, while others predict much heavier candidates, generally known as weakly interacting massive particles (WIMPs), such as the Lightest Supersymmetric Particle (LSP) and others.

WIMPs would have a non-zero scattering cross-section with ordinary matter (thus justifying direct search experiments) and also self-annihilate into standard model particles (indirect detection). These annihilation products would potentially be detectable as an excess in either charged cosmic rays or in photons (X-rays or gamma rays).

Since the WIMPs would have no charge, any annihilation products should be equally split between matter and antimatter. Antideuterons in particular are comparatively rare in the cosmic rays (believed to be entirely of secondary/tertiary origin). Therefore, a clear excess in antideuterons above the flux expected from secondary/tertiary production would be strongly suggestive of dark matter\cite{donato}.

\section{GAPS Science Goals}
\begin{figure}[ht!]
  \centering
    \includegraphics[width=0.7\textwidth]{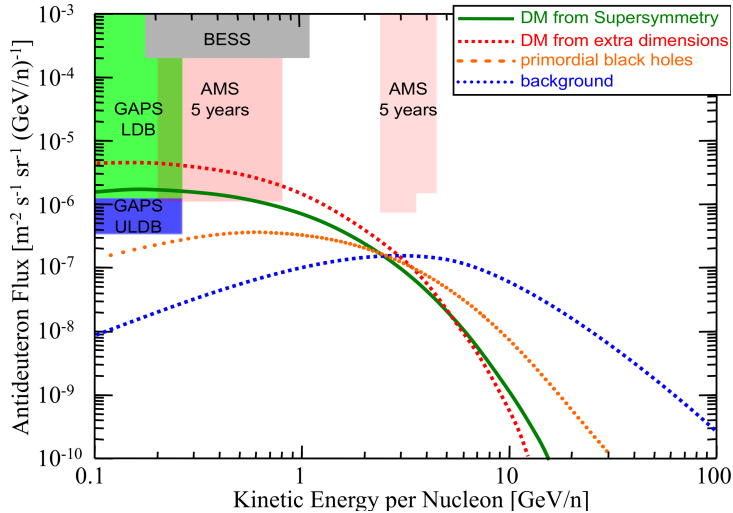}
\caption{Current antideuteron limits from BESS\cite{fuke} shown with  predicted antideuteron fluxes from different dark matter models\cite{baer} and primordial black holes\cite{barrau}, along with expected sensitivities for the operational AMS\cite{doetinchem} and the future GAPS experiments. The expected antideuteron background from secondary production (spallation) is shown in the blue dashed line\cite{duperray}.
}
\label{fig:flux}
\end{figure}

A very promising indirect signature of dark matter are cosmic ray antideuterons. No primary astrophysical sources of antideuterons are expected, and antideuteron  production from cosmic ray spallation in the interstellar medium is expected to be at least five orders of magnitude less efficient than for production of antiprotons\cite{donato}. To date, no cosmic ray antideuterons have been detected, and the experimental upper limits on the antideuteron  flux are well above the antideuteron  flux that might be expected from either dark matter self-annihilation or from spallation (see Fig.\ref{fig:flux}). Experiments with much higher sensitivity are needed.

An essential feature of a possible antideuteron excess from dark matter self-annihilation is the enhanced flux at lower energies (0.1--1 GeV/n) above the expected background from secondary/tertiary production (see Fig.\ref{fig:flux}) In fact, the contribution to the flux due to dark matter could be several orders of magnitude higher than the spallation contribution. Thus an antideuteron  detection in this low energy region could be an essentially background free detection of dark matter. Additionally, the theoretical parameter space probed by this type of search is very complementary to direct detection experiments\cite{baer}.

\section{The GAPS Technique}
It is challenging to discriminate between matter and antimatter particles at high energies. Matter and their antimatter equivalents generally behave the same in calorimeters, scintillators, and other particle detectors. In order to determine a particle’s charge, a strong magnetic field is usually required to deflect the matter and antimatter particles in opposite directions, and these particles must be tracked with good spatial resolution.  This magnetic spectrometer technique has been used in a number of past and present experiments. Magnetic spectrometers have some disadvantages, however, chief among which are a large mass and limited geometric acceptance.

The GAPS technique does not rely on a magnetic field to discriminate particles from antiparticles. Instead, the (negatively charged) antiparticles (antiprotons, antideuterons, antihelium nuclei, etc) are first slowed by a target mass until they can be captured in an atomic orbital of a target atom and form an exotic atom. The exotic atom very promptly deexcites,  emitting X-rays of characteristic energy from the N=8,7, and 6 orbital transitions, and finally annihilates on the nucleus, producing characteristic annihilation products, mostly pions and protons\cite{aramaki}.

The energy of the atomic transition X-rays and the multiplicity of the pion and proton annihilation products are distinct for antiprotons and antideuterons, so the discrimination power of this kind of detector is very high. In order to make an unambiguous detection claim when searching for extremely rare particles such as the antideuteron, it is critically important to be able to suppress backgrounds efficiently, and the GAPS technique has very good rejection potential.

In 2004-2005 a series of GAPS prototype experiments were tested at the KEK accelerator facility in Japan. A number of different target materials were tested and the GAPS exotic atom technique was validated\cite{jcap}\cite{aramaki2}.

In a GAPS science instrument as envisioned here (bGAPS), the target and detector are entirely solid-state. Lithium drifted silicon (Si(Li)) wafers turn out to be a very good choice for the proposed technique. The material offers good X-ray energy resolution, as well as good sensitivity for charged particles, which is useful for tracking both the primary particle and the annihilation products. Also, a large array of Si(Li) detectors would act simultaneously as a degrader to slow down the primary particles and as a target material in which to produce exotic atoms. The only critical requirement for proper operation of the Si(Li) detectors is that they must be cooled to around -35 $^{\circ}$C for good energy resolution.

\begin{figure}[ht!]
  \centering
    \includegraphics[width=0.7\textwidth]{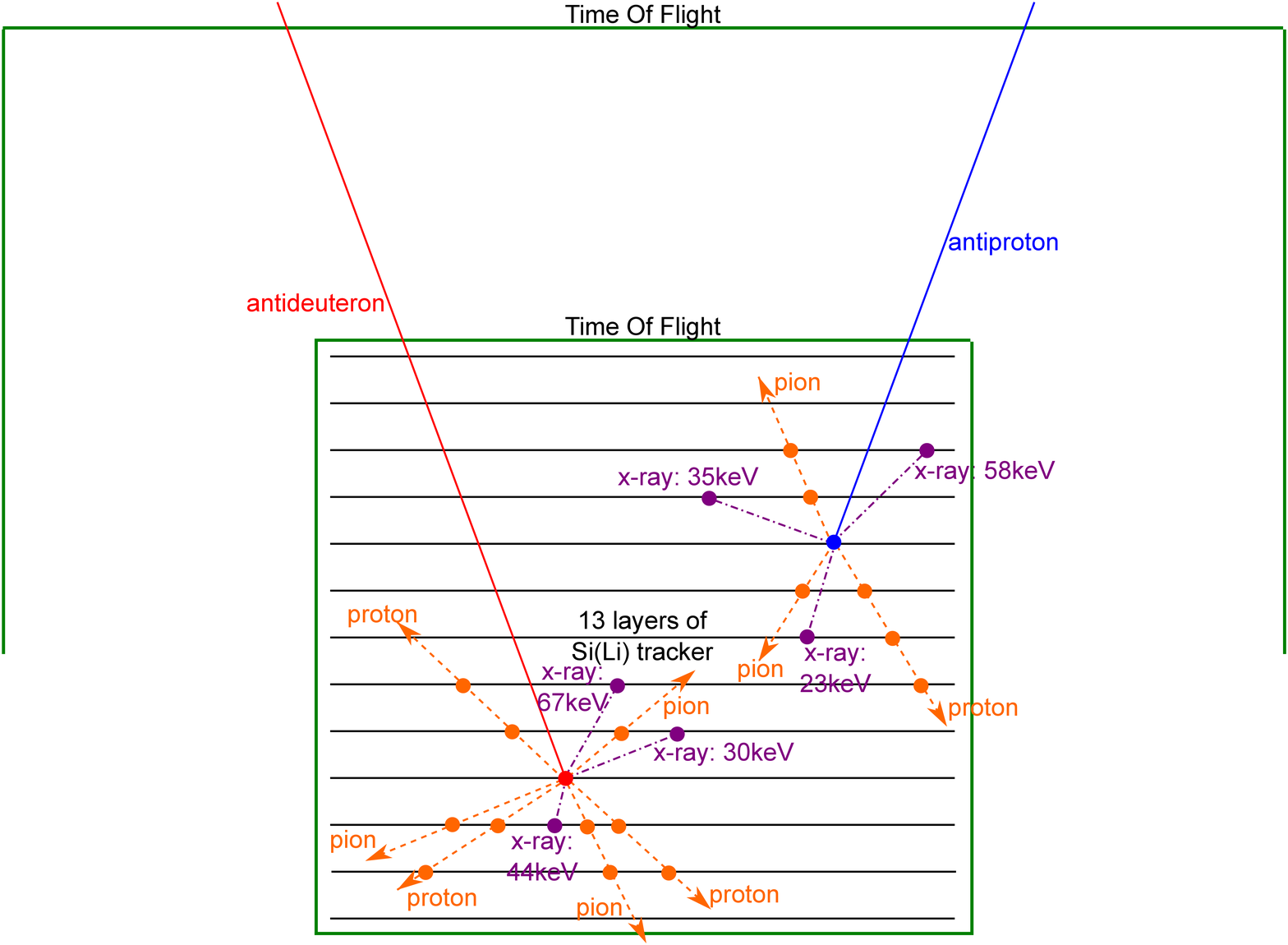}
\caption{bGAPS layout showing the anticipated event structure for an antiproton and antideuteron of the same velocity. The central Si(Li) structure is a ~2 m cube.}
\label{fig:PbarAndDbar}
\end{figure}

In addition to the Si(Li) detector block, bGAPS would require a time-of-flight (TOF) system, which would provide:  a velocity measurement of the incoming primary particle, a charge measurement (to discriminate cosmic ray species with charge Z greater than one), a measure of dE/dx, and a rough trajectory of both the primary particle and any charged annihilation products. Fig. \ref{fig:PbarAndDbar} shows a schematic diagram of the bGAPS concept as well as the event signatures of an antiproton and antideuteron that enter the detector with the same velocity and come to rest within the detector.

\section{The pGAPS Instrument}
A prototype GAPS instrument (pGAPS) has been successfully flown on a high-altitude scientific balloon. The science payload consisted of a time-of-flight system, six cooled Si(Li) detectors, and a fully representative detector cooling system. Since the flight only needed to be of short duration and to contain a representative collection of detectors, minimization of the detector mass or power consumption to the levels required for a full scale GAPS science payload was not needed.

\subsection{Gondola}

\begin{figure}[ht!]
  \centering
    \includegraphics[width=0.7\textwidth]{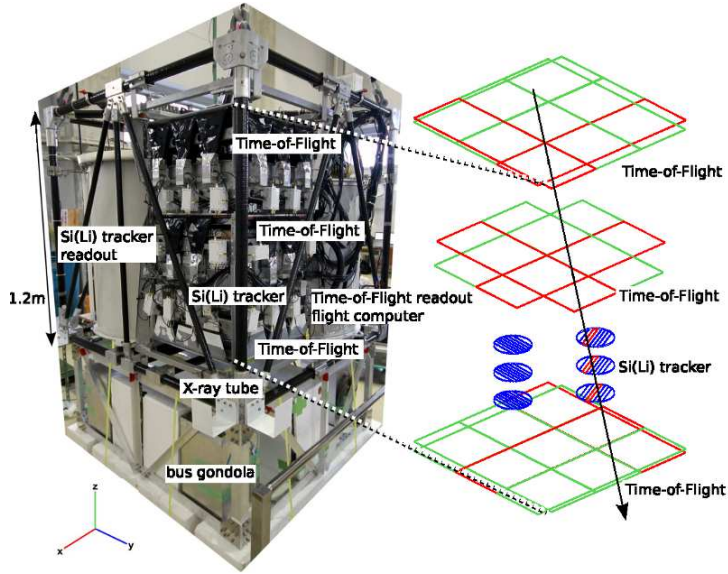}
\caption{Left: Flight ready pGAPS payload without insulation foam. Right: Event display showing the position of the Time-of-Flight and Si(Li) tracker subsystems and a typical clean cosmic ray track reconstructed from flight data.
}
\label{fig:gondolaEvent}
\end{figure}

The pGAPS gondola consisted of two parts: the upper science instrument section and the lower bus gondola.  Fig. \ref{fig:gondolaEvent} shows a photograph of the gondola along with the event display of a typical cosmic ray track from the flight. The total weight of the payload was 510 kg, of which 308 kg was the upper gondola. To reduce the payload mass, the science gondola frame incorporated carbon fiber tube members. The carbon fiber tubes used a custom layered weave matrix to increase the bending moment strength. This construction also added hoop strength at the ends of the tubes to resist crushing. Aluminum connecting blocks were used to join the frame members. Upper and lower horizontal frame members were glued directly into recessed areas in the corner blocks, and steel pins were inserted through the blocks and the tubes for redundant capture. The bus gondola was constructed from aluminum and held the held the telemetry system, batteries, ballast hoppers, and the oscillating heat pipe (Sec.\ref{ohp}).

The gondola was attached to the balloon flight system through mount points at the four corners of the upper frame. The design was modeled in the ANSYS\cite{ansys} finite element analysis package to validate the structure (with populated detector masses) to  10 g  acceleration load  in the vertical direction and a 5 g acceleration load at $45^{\circ}$ from the vertical direction.

Because the balloon was expected to make a water landing, some waterproofing was needed for the readout electronics and the onboard data storage. Additionally, since the TOF readout system and the flight computer were using commercially available modules not designed for operation in vacuum, a pressurized environment was needed to provide convective cooling.

To reduce cost and weight, commercially available high molecular weight polyethylene (HMWPE) plastic vessels were used to make waterproof vessels for the electronics which operated at ambient pressure and a pressure vessel for the TOF electronics and flight computer. Some internal strengthening of the pressure vessel was necessary to prevent deformation on payload descent. Aluminum was used for the tops of the vessels, and cables were fed through these plates. In the case of the pressure vessel, feedthrough connectors were used, while for the waterproof vessel containing Si(Li) readout electronics, cables were run through short lengths of pipe and potted. Lightweight aluminum plates at the bottom of the vessels provided additional strength and solid mount points. 

The readout electronics for the Si(Li) detectors did not require a pressurized environment. But to save them from being damaged in the salt water, they were placed in a custom-designed plastic vessel that was vented to the atmosphere during flight but self-sealing upon landing in water to keep the detectors dry. To close the pipe that vented the vessel during ascent and allowed it to re-pressurize on descent, the vent pipe incorporated a baffle system that contained superabsorbent polymer granules. Upon contact with water, these granules swelled, thus preventing water from entering the vessel. In addition, the venting tube was also fed through a cold box designed to condense inwards diffusing moisture and ensure a dry atmosphere for the Si(Li) detectors. The cold box was kept below water freezing temperature by phase-change-material ice packs. The plastic vessel that housed the Si(Li) detectors also used the same waterproof baffle system.

The main power distribution system was mounted on an aluminum backplane and consisted of a power distribution box and two boxes of DC-DC converters (Sec. \ref{dcdc}).

\subsection{Attitude Control System}

\begin{figure}[ht!]
  \centering
    \includegraphics[width=0.7\textwidth]{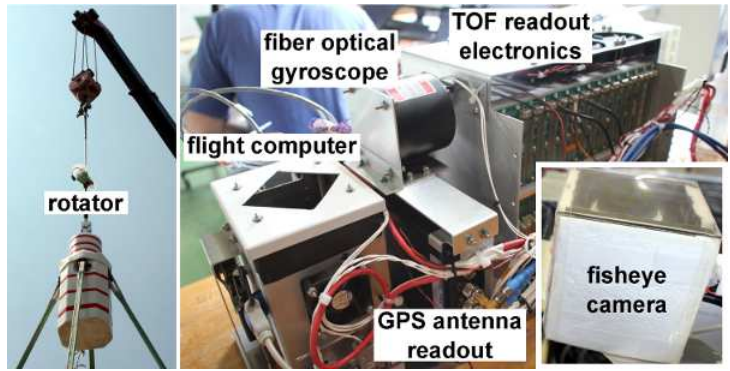}
\caption{Components of the attitude control system.
}
\label{fig:rotator}
\end{figure}

The attitude control system (ACS), shown in Fig. \ref{fig:rotator}, consisted of several redundant  systems. Active pointing was provided by a rotator mounted between the payload and balloon and  controlled with a voltage from the flight computer. The goal during the flight was to have the cooling system space radiator always pointing away from the sun. The pointing direction could be determined by three different redundant methods. The primary method was a global positioning system (GPS) antenna, the secondary method was a fiber-optic gyroscope, and the tertiary method was a fisheye camera placed on the top of the gondola with the balloon and the sun in its field of view. 

The rotator was designed to be powerful enough to rotate the gondola with respect to the balloon at float altitudes of ${\sim}$30 km. Angular information from the GPS antenna or the fiber-optic gyroscope served as a feedback to the flight software and was evaluated at a rate of ${\sim}$5 Hz. The underlying control law was based on moment-of-inertia simulations and measurements of eigenfrequencies of the system. It was also possible to select different sets of control law parameters, or to select the pointing angle manually during flight. 

\begin{figure}[ht!]
  \centering
    \includegraphics[width=0.7\textwidth]{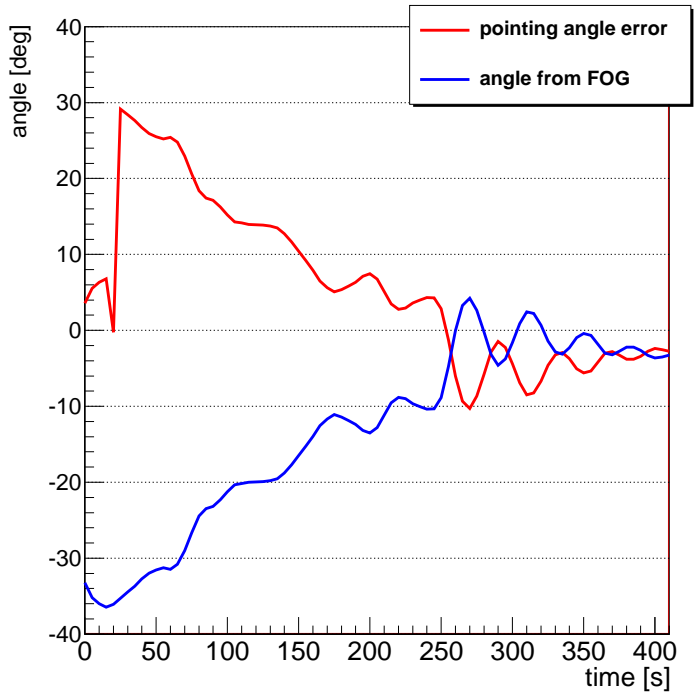}
\caption{Ground testing of attitude control system showing payload pointing recovery from an initial ${\sim}30^{\circ}$ displacement. The instrument pointing angle was measured by the fiber-optic gyroscope. 
}
\label{fig:rotatorWork}
\end{figure}

The ACS was tested with the whole payload attached as part of ground testing. The gondola was suspended and a specialized parameter set for the test setup was used. Fig. \ref{fig:rotatorWork} shows the pointing error and the actual angular pointing measured with the fiber-optic gyroscope during a test with the payload recovering from a ${\sim}30^{\circ}$ initial displacement. This demonstrates that the rotator was able to rotate the gondola successfully to a certain set point and stabilize it there to within a few degrees.  Unfortunately, due to an operational mistake (applying -5 V control voltage directly after +5 V) which created too much torque and broke the internal mechanism, no pointing was available during the flight. As discussed later in the paper however, this did not adversely impact the completion of the flight goals.

\subsection{Si(Li) Detectors}
The pGAPS payload carried six commercially available\cite{semikon}, circular, lithium-drifted silicon detectors, each with an active diameter of 9.4 cm. Five Si(Li) detectors were 2.5 mm thick, and one was  4.2 mm  thick. The upper (p+) surface was segmented into eight strips and had implanted boron contacts. The lower (n+) side was not segmented and had lithium contacts. The nominal operating temperature for the Si(Li) detectors was specified at around -35 $^{\circ}$C with an operating voltage of around 200 V. 

The goal for the X-ray resolution is 3 keV full width at half maximum (FWHM) at 60 keV and is driven by the need to discriminate between exotic atoms formed by antideuterons and antiprotons. In laboratory testing with conventional bench-top electronics and an Am-241 X-ray source, most channels easily achieved the design resolution.

With the detectors integrated into the payload, the average resolution was 5.6 keV FWHM for X-rays from an externally positioned Am-241 source. This figure reflected both the inherent broadening of the X-ray line due to scattering within the instrument structure, some increased electrical noise from other payload subsystems (cooling pump, nearby temperature sensors, TOF, etc), and the response of the readout electronics. These readout electronics were repurposed from the NCT\cite{nct} payload and thus not fully optimized for the pGAPS detectors. Also, for the full GAPS payload the exotic atom creation will happen inside the tracker volume and therefore the transition X-rays will be detected without passing through much passive material.

\begin{figure}[ht!]
  \centering
  \includegraphics[width=0.7\textwidth]{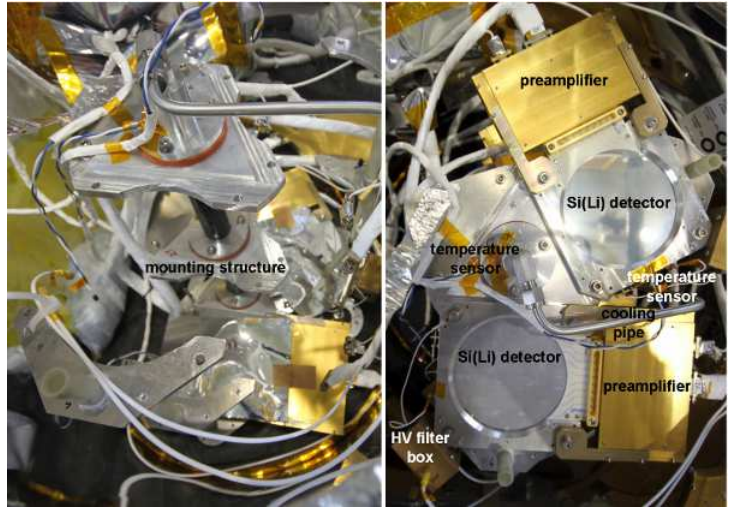}
	\caption{Left: Si(Li) detector mounting structure. Right: Mounted detectors. For scale, each Si(Li) detector was 9.4 cm in diameter.}
\label{fig:siliHome}
\end{figure}

The detectors were mounted on a tree-like support structure inside an HMWPE plastic vessel (Fig. \ref{fig:siliHome}). A preamplifier module was mounted directly adjacent to each detector. Also, each high-voltage line powering the detectors had a passive low-pass filter to reduce electronic noise. The central support rod contained the cooling pipe, through which fluid cooled down the mounting structures. The six detectors were arranged in two columns of three detectors, with each layer separated by 20 cm. 

Electrical isolation of the detectors was necessary, so inside the detector vessel a Faraday cage was formed by taping 127 $\mu$m thick aluminized boPET foil to the inside vessel walls with the plastic side facing the inside. The boPET shield was electrically tied to the internal cooling pipe and mounting structure inside the detector vessel, which was taken as the common ground. All feedthroughs and the joint between the detector vessel lid were sealed electrically and tied to the central grounding structure. The central cooling pipe was electrically isolated from the outside piping system by a segment of G10 (glass reinforced epoxy laminate) piping to avoid external pick-up noise. 

After prolonged operation of the Si(Li) detectors on the ground it was discovered that they were quite robust and that passivation was not necessary. All that was needed for ground operation was a simple purging system to keep the vessel filled with dry nitrogen to prevent condensation. The vessel was sealed well enough to maintain a dry atmosphere, and  diffusion air into the vessel was not an important effect. The vessel also protected the detectors during the water landing. During flight, the vessel was vented to ambient atmosphere, but with the baffle system containing polymer granules as described above. 

\subsection{Cooling system}
Since the Si(Li) detectors must be operated between -25 $^{\circ}$C to -35 $^{\circ}$C, a fully flight representative cooling system was flown. While not essential for the operation of the detectors on such a short flight, this was crucial for validating the cooling approach and the thermal model needed for development of a GAPS science payload. 

\begin{figure}
\centering
\begin{minipage}{0.49\textwidth}
  \centering
  \includegraphics[width=1.0\linewidth]{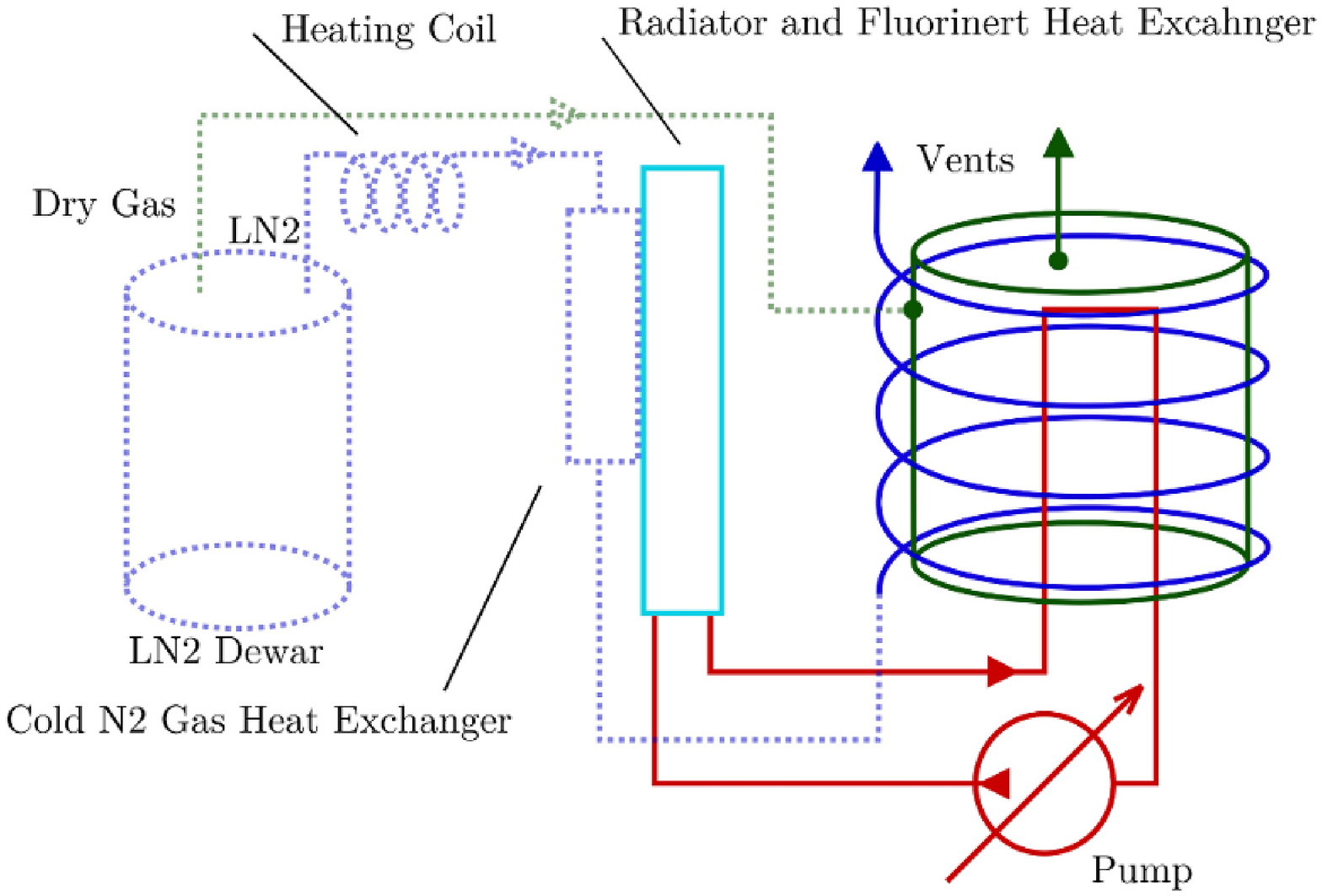}
  \caption{Simplified schematic of the pGAPS cooling system. Dashed components are for ground operation only (not used in flight).}
  \label{fig:cooling}
\end{minipage}\hfill
\begin{minipage}{0.49\textwidth}
  \centering
  \includegraphics[width=1.0\linewidth]{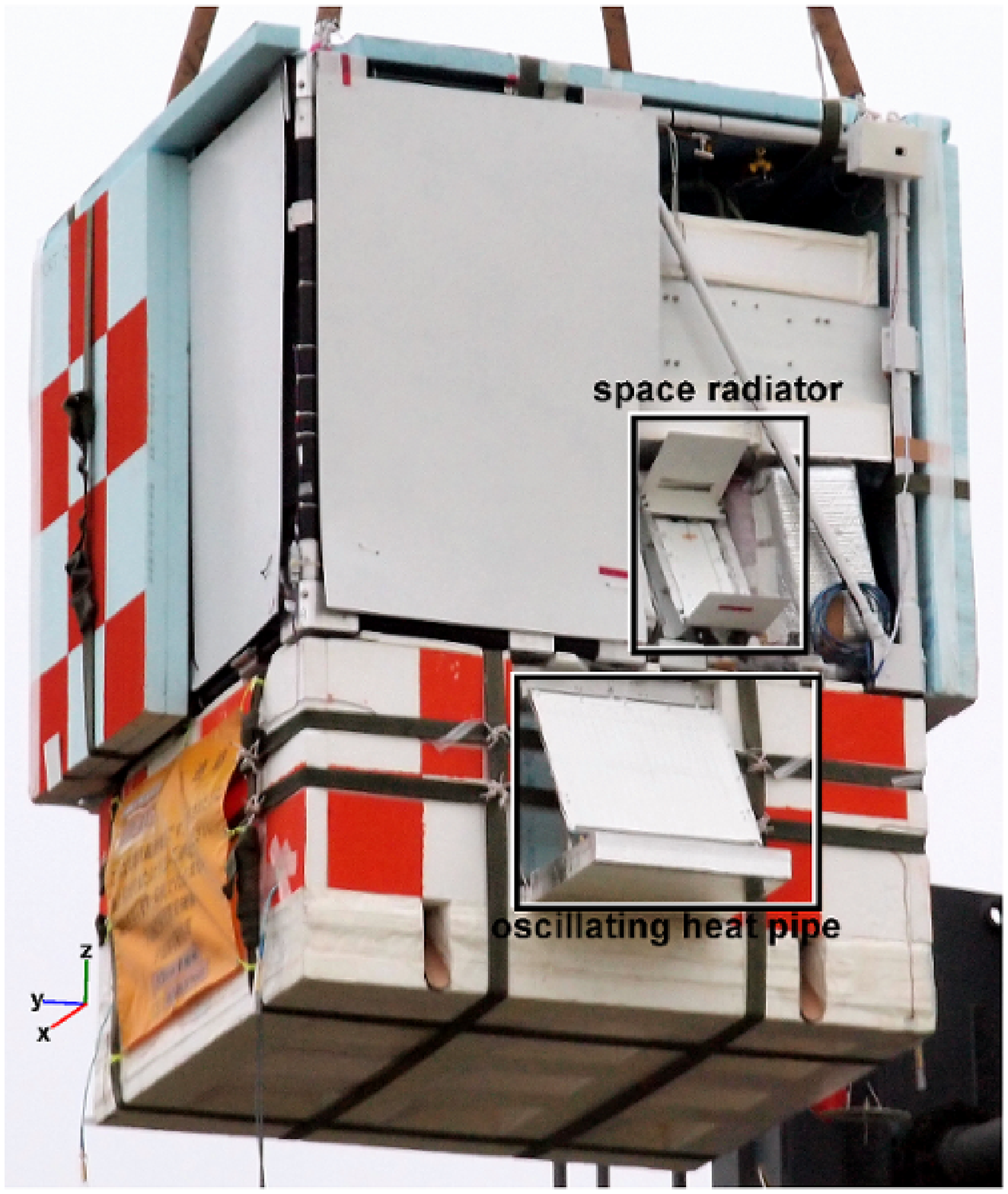}
  \caption{Space radiator and oscillating heat pipe shown on the (flight ready) pGAPS payload.
}
  \label{fig:launchReady}
\end{minipage}
\end{figure}

The cooling system used on pGAPS contained a radiator, a closed-loop coolant path with Fluorinert fluid, and a pump. This system is shown schematically in Fig. \ref{fig:cooling}. The cooling pipe cooled down the mounting structure of the Si(Li) detectors (Fig. \ref{fig:siliHome}, left). During flight operations, the radiator would face the anti-sun side of the instrument and serve as a heat dump (see Fig. \ref{fig:launchReady}). The pump speed could be set from the ground to regulate the rate of heat transfer from the detectors to the radiator. For flight,  aluminum shields were placed above and below the radiator to block solar albedo and to further decouple the radiator from the instrument.

For ground operation, a heat exchanger was mounted to the radiator and cooled with cold nitrogen gas, thus using the radiator and coolant system in a flight-like way in all ground tests. Cold, gaseous nitrogen exiting the heat exchanger was also passed around the outside of the Si(Li) detector vessel in a coil of copper tubing to cool down the detector vessel more quickly. In addition, it was also possible to direct cold nitrogen purging gas into the detector vessel. The cooling strategy during ground testing thus utilized a mix of conductive and radiative cooling. It would not have been possible to provide enough cooling power through either conductive or radiative cooling alone during ground operations.

As mentioned previously, the rotator for the gondola failed, so no pointing control was available.  At float the gondola had a natural rotation period of ~5 minutes. Only limited cooling of the detectors was provided by the active cooling system. Ground pre-cooling and thermal inertia provided the rest.  At the time of the launch the mean detector temperature was -35 $^{\circ}$C and all detectors stayed below -15 $^{\circ}$C until the end of the flight. The active cooling system was operated sufficiently for a full validation of the cooling system design and for full validation of the thermal model.

\subsection{Passive Thermal Control}
Aside from the Si(Li) subsystem that had active thermal control (cooling), the rest of the payload relied on passive techniques to maintain operating temperatures. Generally this meant keeping all electronics between -10 $^{\circ}$C and +40 $^{\circ}$C. Another important consideration was to prevent heat from solar radiation, solar albedo, or from other subsystems from reaching the Si(Li) detector vessel or radiator. 

The payload was partially enclosed in 0.02 g/cc foam insulation (between 5--10 cm in thickness) to shield the sides (and top) of the payload that would be facing direct sunlight. Additionally, the top and sun-facing sides of the instrument were also covered in a layer of 127 $\mu$m aluminized polyester. Part of the anti-sun side and half of the port side were covered in 0.8 mm (white painted) aluminum sheets. This shield protected the pressure vessel from solar albedo and direct sunlight, in the event that the instrument rotated, while allowing the pressure vessel to radiate some of its (${\sim}$300 W) heat to space. The power converter boxes (and several other electronics boxes) were mounted to the rear of the payload on a heavy (painted) aluminum plate that served as a heat sink and passive radiator. 

The top (science) section of the payload was separated from the bottom (bus) section by 5 cm of styrofoam to isolate the two sections from each other thermally. Other subsystems, such as electronics boxes, were enclosed in a combination of styrofoam, aluminized bubble wrap, and aluminum tape. The detector vessel was surrounded with aluminized bubble wrap, which served as a barrier to heat from other gondola subsystems.

\subsection{X-ray Calibration System}
In order to provide a calibrated source of X-rays throughout the flight, a commercially available X-ray tube (40kV MAGNUM® from MOXTEK\cite{moxtek}) was flown. The tube used a 0.25 mm silver target to produce X-ray fluorescence lines appropriate for calibrating the Si(Li) detectors. The aluminum housing and silicone oil bath attenuated the radiation emitted inside the tube to safe levels. After attenuation the main X-ray features remaining were the silver K-beta line at 26 keV and a broad emission centered near 36 keV. The silicone oil in which the X-ray tube was immersed also prevented high voltage (HV) breakdown.

\begin{figure}
  \centering
  \includegraphics[width=0.7\textwidth]{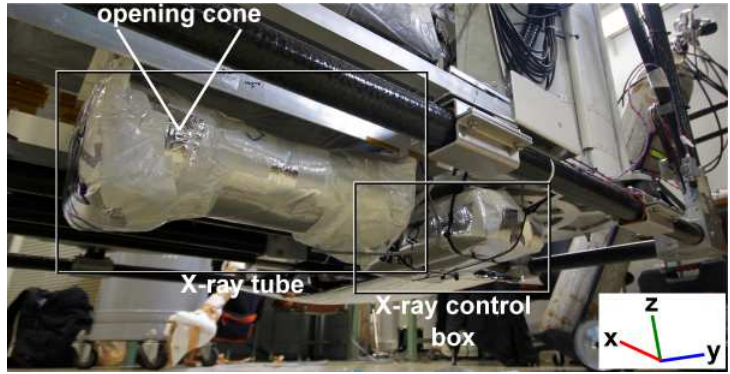}
	\caption{The X-ray system mounted below the Si(Li) detector vessel. The X-ray tube and control box are wrapped in aluminized bubble-wrap and tape for thermal insulation.}
\label{fig:xraymount}
\end{figure}

One of the design requirements for the X-ray system was that it should be capable of illuminating all Si(Li) detectors with a flux sufficient to produce ${\sim}$1000 counts in each Si(Li) strip over the course of each three minute long calibration interval, while not exceeding a maximum trigger rate of ${\sim}$2 kHz in the most illuminated detector strips (limited by readout electronics). The mounting position and absorption material were optimized to meet these flux requirements (Fig. \ref{fig:xraymount}). The tube was operated every half-hour during the flight during which the Si(Li) subsystem was operated in self-trigger mode.  

\subsection{Time-of-Flight System}		
\begin{figure}
  \centering
  \includegraphics[width=1.0\linewidth]{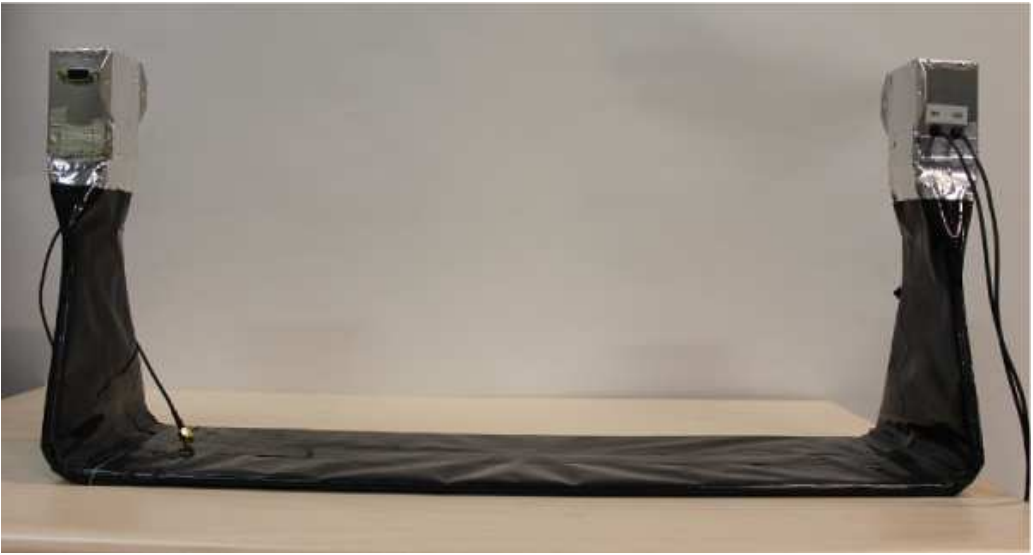}
\caption{Photograph of one pGAPS scintillator paddle. The paddle includes a 50 cm long BC-408 plastic scintillator slab, two acrylic light guides, a light-tight wrapping, and fast PMT assemblies mounted on each end.
}
\label{fig:paddleandtube}
\end{figure}

The pGAPS Time-of-Flight system (TOF) consisted of three layers of crossed plastic-scintillator paddles (see Fig. \ref{fig:paddleandtube}). Each paddle was made from a 15 cm $\times$ 50 cm piece of 3 mm thick BC-408\cite{bc408} polyvinyltoluene (PVT) based scintillator. Each end of the scintillator was attached to a curved, acrylic light guide coupled to a fast photomultiplier tube (PMT). Hamamatsu R7600-200 compact 18$\times$18 mm$^2$ Ultra-Bialkali PMTs were used\cite{r7600}, which offered exelent gain and timing characteristic, in a very compact package.

All paddle assemblies were wrapped with aluminum foil to provide specular reflection (at a randomized reflection angle) for any light that escaped from the scintillator or light guides. By randomizing the reflection angle, some percentage of the light which exited the plastic optics would be reflected back at a smaller angle sufficient to be recaptured by total internal reflection. Computer simulations (discussed in more detail in section \ref{tofsim})  and laboratory studies with test paddles were conducted to compare aluminum foil to several other candidate wrapping materials (white Teflon, white Tyvek and aluminized boPET). Both simulations and experiments confirmed that the foil provided a slightly higher total light collection efficiency. Good agreement with the simulation code and the measured detector performance was achieved, thus validating this tool for future use in bGAPS design.

Light-tightening of the paddles was done with a layer of Delta 6 mil (0.15 mm) dark-room blackout material (plastic sheet with carbon black).  Further light-tightening around seems (joints between the blackout material and the PMTs, etc) was achieved with adhesive-backed aluminum tape.

\begin{figure}
  \centering
  \includegraphics[width=0.7\textwidth]{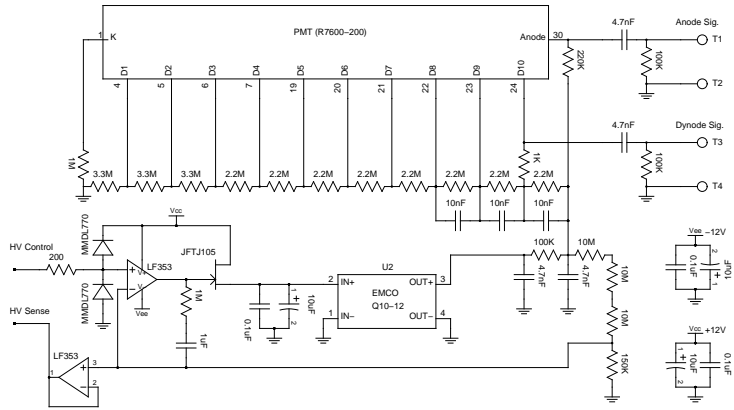}
	\caption{Schematic of the pGAPS TOF PMT base circuit. Each base contained a dedicated HV supply, so no high voltage needed to be distributed outside of the potted base assemblies. The operating voltage was also programmable with a 0-5 V control voltage for each individual PMT.}
\label{fig:pGAPSbaseV2nim}
\end{figure}

A custom PMT base circuit was used for safe operation in partial vacuum (Fig. \ref{fig:pGAPSbaseV2nim}). To prevent high voltage (HV) breakdown, the PMTs and bases were potted with encapsulant and positive HV was used so the photocathode would not be at high potential (and thus cause corona discharge).  Furthermore, integrating the HV supply for the circuit into the base eliminated the need  to distribute HV externally. Finally, the first dynode needed to be tapped and read out in addition to the anode, which was not a function offered by the stock Hamamatsu base.

HV was provided internally to each base circuit by the Q10-12 compact switching supply from EMCO\cite{emco} with regulation added by putting the HV supply in the feedback loop of an operational amplifier. This provided output voltage stability better than $\pm1\%$. Without regulation the HV output varied by ${\sim}20\%$ over the full operating range expected in flight (-40 $^{\circ}$C to +40 $^{\circ}$C). 

Each PMT and base circuit were potted in a thin-walled aluminum housing with welded seams. Stycast 4640 White\cite{stycast} was used for potting, which offers excellent dielectric breakdown and flow characteristics. The potting compound was vacuum degassed both prior to, and after, pouring into the housings. The PMT assemblies were optically coupled to the light guides with RTV615 optical RTV\cite{rtv615}. Additional strain relief was provided by tabs which extended down from the base housing and clamped around the light guide end.

The typical photoelectron count in each PMT was ${\sim}15$ per minimum ionizing particle, as measured during ground testing with atmospheric muons. This figure took into account the quantum efficiency (QE) of the PMTs, so the actual photon count incident at each PMT would have been closer to 25-35 (the peak QE was ~43\% with the Ultra Bialkali tubes used).

\begin{figure}
  \centering
  \includegraphics[width=0.7\textwidth]{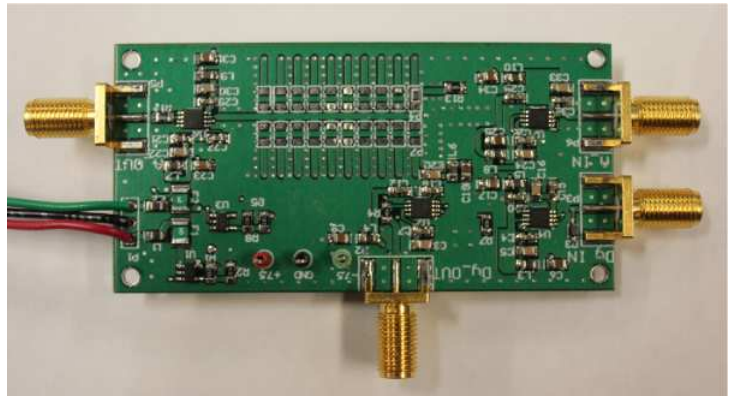}
	\caption{The custom preamplifier board used for the pGAPS TOF using THS3201 current feedback operational amplifier ICs. A Rhombus Industries SP24A-1005G analog delay line was used to delay the anode pulses by 100 ns. (Delay IC on back of board.)}
\label{fig:preamp}
\end{figure}

A custom preamplifier board (Fig. \ref{fig:preamp}) was mounted close to each PMT assembly. THS3201 current feedback operational amplifiers were used, which offered very wide bandwidth and low power operation. The preamplifier board also introduced a 100 ns delay to the anode pulses using a delay line chip from Rhombus Industries (the SP24A-1005G). This delay was needed so that a trigger could be formed (from the dynode pulses) in the TOF trigger logic in time to instruct the analog-to-digital converters (ADCs) (used for anode readout) to begin integrating. The preamp also inverted the (positive) dynode pulse to match the (negative) polarity the time-to-digital converters (TDCs) (section \ref{trs}) were expecting.

\subsection{Oscillating Heat Pipe (OHP) Test}
\label{ohp}
The Oscillating Heat Pipe\cite{ohp} offers several potential advantages over the proven active pumped-Fluorinert cooling system described above. Chief among these are a reduced power consumption and increased mechanical simplicity.

\begin{figure}
  \centering
  \includegraphics[width=0.7\textwidth]{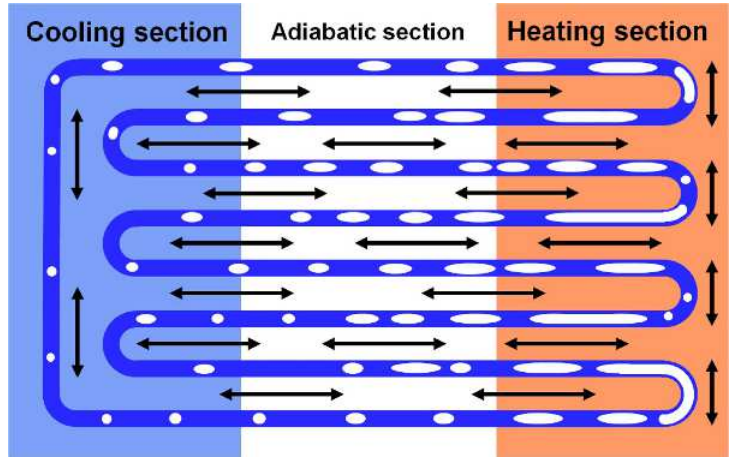}
	\caption{A conceptual diagram of the OHP operating principle.}
\label{fig:ohpDig}
\end{figure}

The operating principle of an OHP is shown in Fig. \ref{fig:ohpDig}. A capillary tube containing a phase-changing working fluid (R-410A refrigerant in this case) threads back and forth between a cooling section (the radiator) and a heating section. Small vapor bubbles form in the working fluid, then expand in the heating section, and contract in the cooling section. Driven by the collapse and expansion of the vapor bubbles, pressure and temperature fluctuations quickly set up thermo-hydrodynamic waves, which drive the fluid back and forth between the hot and cold sections in a completely passive process.

\begin{figure}
\centering
\begin{minipage}{0.5\textwidth}
  \centering
  \includegraphics[width=0.9\linewidth]{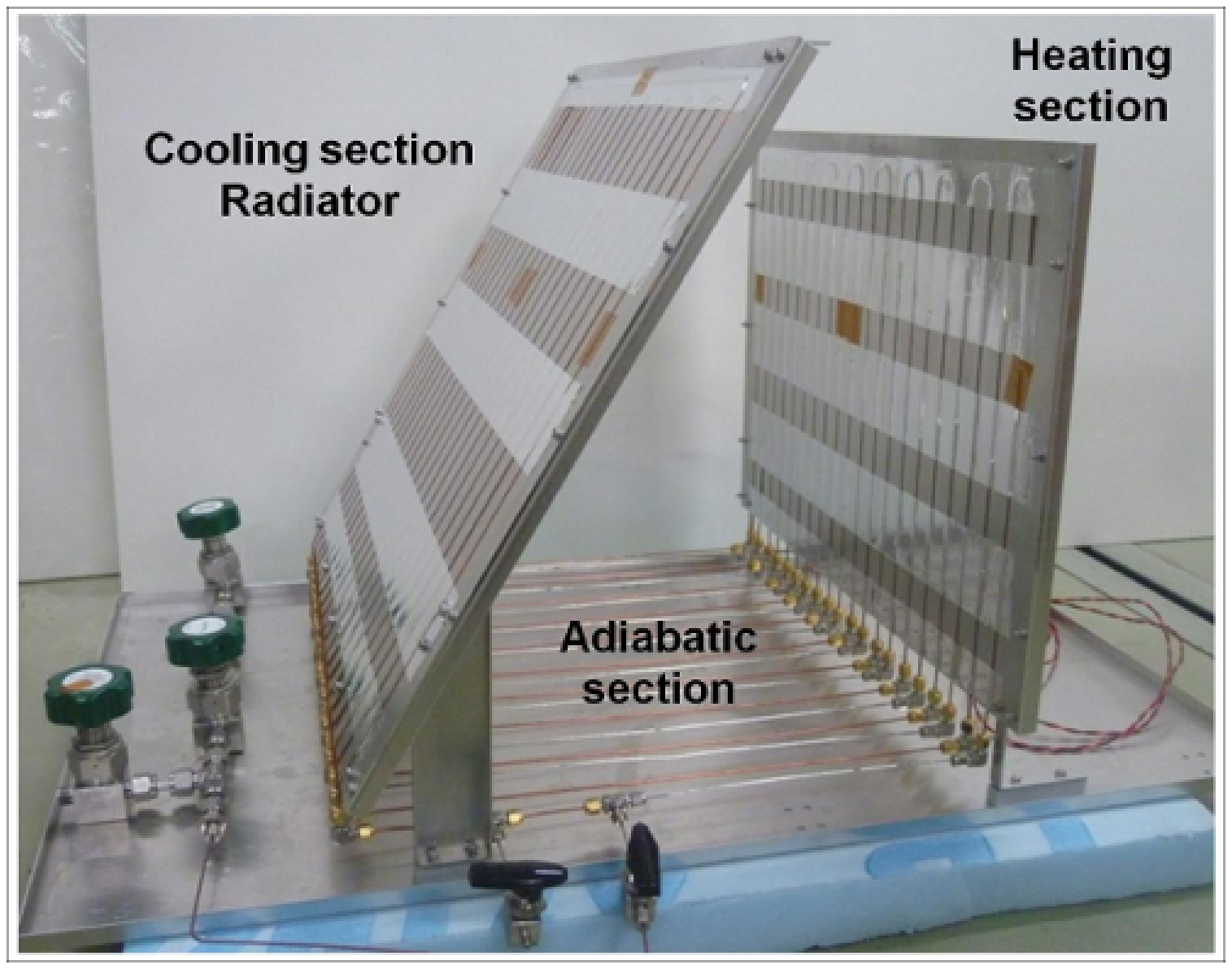}
\end{minipage}%
\begin{minipage}{0.5\textwidth}
  \centering
  \includegraphics[width=0.9\linewidth]{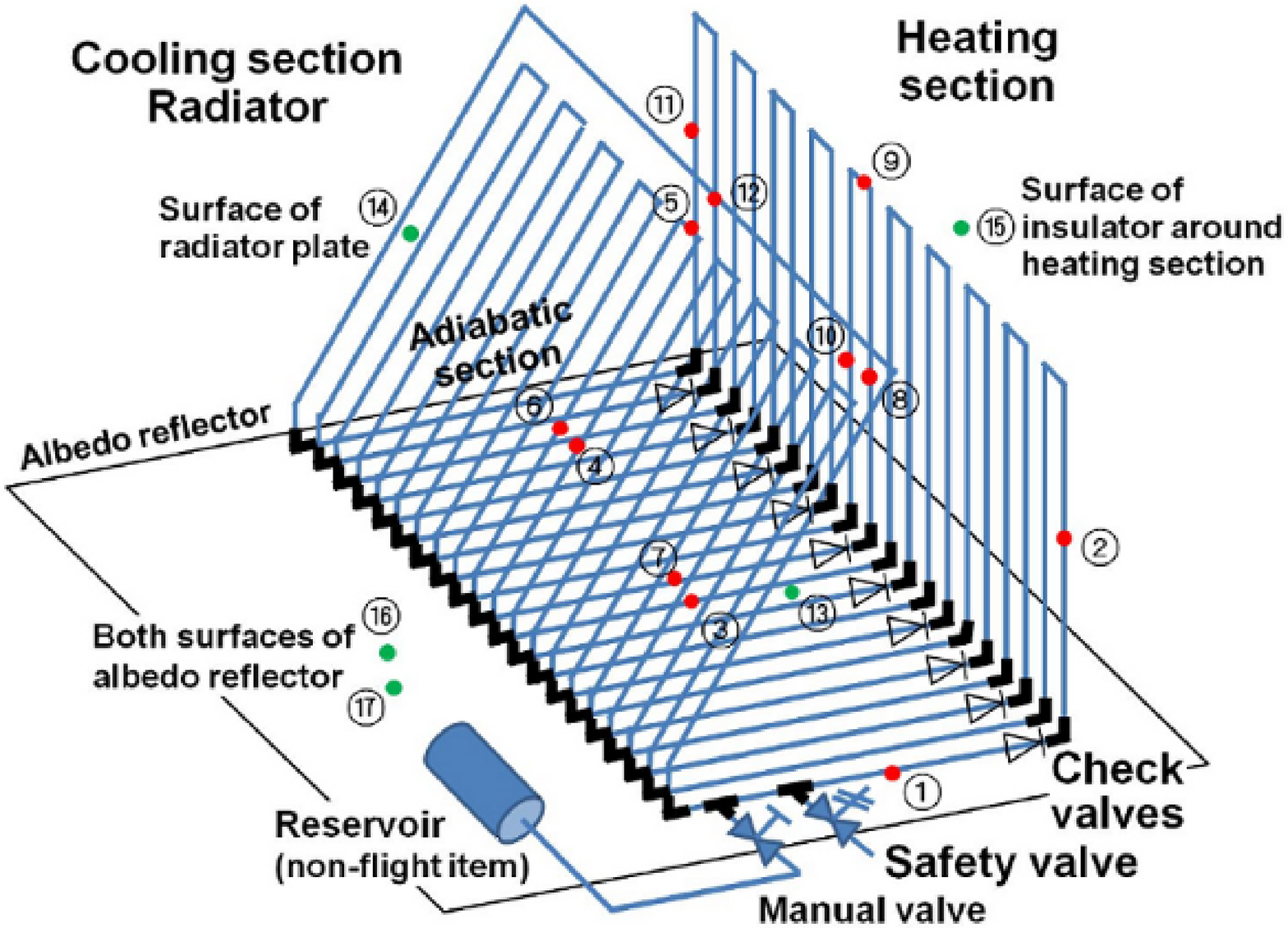}
\end{minipage}
\caption{Left: A photo of the pGAPS OHP before the thermal insulation was applied. Right: A diagram of the pGAPS OHP. The numbers 1--17 indicate locations where temperatures were measured by thermocouples. }
\label{fig:ohpFigs}
\end{figure}

The OHP is a radical new concept for a cooling system and the pGAPS flight is the first operation of an OHP on a balloon payload (Fig. \ref{fig:ohpFigs})\cite{ohpgaps}. This is also the first operation of an OHP at temperatures as low as -40 $^{\circ}$C.

In order to obtain very clean validation data, the OHP was thermally isolated from the rest of the gondola. A portable data-logger was used to record all in-flight readings from the OHP system and no data transfer was available during flight. The recorded data were successfully recovered after the flight. 

The prototype OHP was mounted on the anti-sun side of the payload. Heat was supplied by two 7.5 W heaters, and radiated to space through a plate at the cold end. The loop length was ${\sim}$1 m, and thus this OHP system represents a scaled-down version of the OHP system that would be required for bGAPS. 

\subsection{Electrical System Overview}
\label{dcdc}
Raw power during the flight was provided by a large array of D-cell Lithium batteries contained in the lower gondola section. The array was wired in series and parallel modules to provide ${\sim}$30 V with sufficient stored energy for the entire flight and several ground tests (hang and compatibility). Under nominal data taking conditions, the system had a power consumption of 430 W. For normal ground operations, an AC-powered bench supply provided raw power to the the payload. The raw battery or external power were converted and regulated by Vicor VI series DC-DC converter modules (and Vicor ripple attenuation modules) to supply all needed voltages to payload subsystems.

The output of each DC-DC converter could be turned on or off by commands from the flight computer using discrete I/O operations explained in section \ref{fc}. The power distribution box (and the DC-DC converter boxes) monitored current draw and supply voltages (and were reported by the housekeeping system). In addition, the power distribution box was equipped with a marine switch for quick power shutdown upon payload recovery at sea. The X-ray tube had three redundant means of powering down: by command from the flight computer, by having its power cables cut by a rope cutter powered by a seawater battery immediately upon landing, and finally by the marine switch mentioned above.

High voltage for the TOF PMTs was provided locally inside each PMT base assembly. The voltages for each PMT were individually settable by adjusting a 0--5 V control line going to each base. The HVs were also continuously monitored by the TOF housekeeping system, using a low voltage readback from each base which was proportional to the applied HV.

The voltages required by the Si(Li) detectors were between 185 V and 240 V, so no corona discharge precautions were needed. Each detector was powered by a separate HV supply, which was mounted in the tracker electronics rack. These supplies were pre-adjusted on the ground to voltage levels optimized for each detector and were not modified in flight. However, the supplies could be commanded to ramp up or down to their set voltage levels. The ramping circuit was included  to avoid damage from electrical breakdown.

The grounding scheme of the payload followed a strict isolation strategy of the different subsystems to avoid ground loops and to prevent electrical noise from spreading among the different systems. As explained above, special care was taken to have a well-defined ground for the noise sensitive Si(Li) detector modules to ensure good X-ray resolution. Also, the motor of the pump was electrically isolated from the cooling-loop piping. 

\subsection{TOF Readout System}
\label{trs}
\begin{figure}
  \centering
  \includegraphics[width=0.7\textwidth]{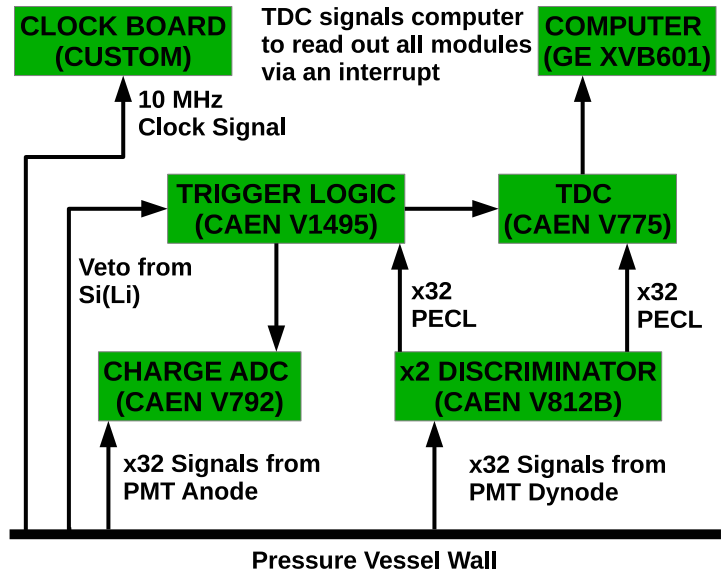}
	\caption{Schematic diagram of the TOF readout electronics. Detailed operation of the system is described in the text.
}
\label{fig:flight_config}
\end{figure}

The TOF data acquisition system used a commercially available VME crate, backplane and modules (shown schematically in Fig. \ref{fig:flight_config}). The crate utilized the custom built DC-DC power converters discussed above for the +5 V, +12 V, and -12 V supply lines. Additional decoupling capacitors were attached to the VME backplane to stabilize voltages after the ${\sim}$3 m of cable.

The PMT dynode signals went to programmable, leading-edge discriminator modules (Caen V812B) that produced two digital outputs. One discriminator output went to a CAEN 1495 logic module, based on a field-programmable gate array (FPGA), that read the discriminated pulses to generate a trigger in a three-step process. First, a paddle trigger was formed by a logical OR of the PMT pulse from either end of a paddle that exceeded the discriminator threshold. A coincidence of two paddle triggers from perpendicular paddles of the middle layer (see Fig. \ref{fig:gondolaEvent}) caused a middle-layer trigger. Finally, a middle-layer trigger coincident with any individual paddle trigger from the top or bottom layers generated a system trigger. The logic module distributed the system trigger to the time-to-digital converter (TDC), analog-to-digital converter (ADC), a clock board (described below), and to the Si(Li) system. It also received a busy signal from the Si(Li) system that would veto all further triggers that arrived while the Si(Li) system was being read out.

The other discriminator output went to a 32-channel, 12-bit TDC module (Caen V775) to provide timing measurements. The TDC operated in common stop mode with a time step of 50 ps per digital channel count. When the TDC received a trigger from the logic module, it generated an interrupt that signaled the CPU to read out the TDC, ADC, and clock board. The PMT anode signals were delayed by 100 ns in the preamps and sent to a Caen V792 32-channel, 12-bit charge integrating analog-to-digital converter (ADC).

A custom-built, FPGA-based, digital VME board read the 10 MHz common clock used in the payload to synchronize event numbers between the different subsystems. The counter was 32-bit and would thus roll over every 429.5 seconds. A SYNC line was also present to synchronize all sub-system event numbers (either on startup or if mismatches were observed).

Three microcontroller-based data acquisition boards (LabJack U6 OEM version) with USB interface performed HV control and readback. These boards had 14 analog inputs for housekeeping and 20 general-purpose digital I/O lines that could be attached to custom daughter boards. A custom daughter board was used that provided 16 digital-to-analog converter (DAC) lines (12-bit), settable from 0 to 5 V, to control the HV of the TOF PMT bases. Of the 42 analog inputs, 32 were used for HV readback, while the remaining 10 inputs monitored temperatures throughout the payload. 

The XVB601 single board VME computer (manufactured by the General Electric Company) running Fedora Core 12 controlled the VME crate and read out the data. A fully-threaded C++ data acquisition program performed the following functions: the program 1) received and executed commands sent from the flight computer, 2) periodically read out the HV and temperature data, 3) read out and stored the TDC, ADC, and clock module data upon receiving an interrupt from the TDC module, and 4) sent any new data back to the flight computer (for transmission to ground control) and recorded a copy of all the data locally on an external 8 GB USB drive. The USB drive allowed for prompt recovery of the flight data without having to recover and start up any of the payload computers after the (water) recovery. The TOF computer communicated with the flight computer via a standard Ethernet switch using the UDP protocol.

\subsection{Si(Li) Readout System}
\begin{figure}
  \centering
  \includegraphics[width=0.7\textwidth]{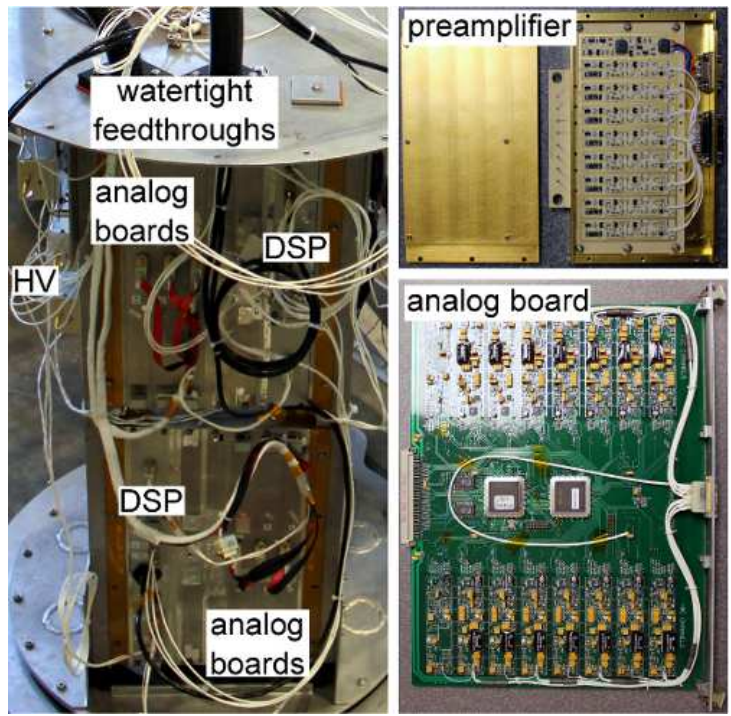}
	\caption{Si(Li) tracker readout electronics. (left) the card-cage readout racks. (right, top), one of the preamplifier boards and housings. (right, bottom), one of the analog readout boards.
}
\label{fig:siliReadout}
\end{figure}

The positive going signal from each strip of the Si(Li) detectors was amplified by a factor of 30 in a preamplifier module mounted adjacent to each detector (Fig. \ref{fig:siliReadout}, right, top). The preamplifier boards were housed in gold-plated aluminium boxes to provide good electrical shielding. The operation of one preamplifier for eight strips required +6 V and -6 V and consumed about 400 mW of power in total. The average noise level for all preamplifier channels was $1.2\pm 0.6$ keV and therefore only a minor contribution to the required 3 keV FWHM resolution for 60 keV X-rays. The signals were transmitted via flexible, coaxial ribbon cables to the analog and digital readout processing in the tracker readout vessel. Two card cages held three signal processing boards each, one for each detector and also provided the voltages for the preamplifiers. The boards connected to a common backplane, which transmitted the signals to the digital signal processing unit (DSP). The card cages and boards had previously flown on the NCT experiment\cite{nct} and were modified to match the needs of GAPS.

Each signal processing board (Fig. \ref{fig:siliReadout}, right, bottom) read out one Si(Li) detector (eight strips) and had both eight high-gain and eight low-gain channels. The incoming tail pulse from the preamplifier was simultaneously read out by a high-gain channel (with an amplification factor of 20) and a low-gain channel (with a damping factor of 0.5). The two gain ranges were required as the tracker electronics needed to deliver both good X-ray resolution in the range of 10--100 keV and measure the energy deposit of charged particles (0.5--10 MeV). 

For both signal branches, the amplification stage was followed by a Gaussian shaper and compared to a low-level discriminator (LLD), an upper-level discriminator (ULD), and tested for being negative. The number of counts above the different discriminator thresholds were recorded by a central FPGA. The trigger for the analog-to-digital (A/D) conversion was generated if the signal was non-negative, above the LLD threshold, as well as below the ULD threshold. This trigger was forwarded to an additional FPGA, the so-called state FPGA, and the A/D conversion was executed if the additional requirement of a peak detection from the shaped pulse was fulfilled. The digitized energy depositions were stored in a FIFO. The individual LLD thresholds for each channel were controlled via registers on the state FPGAs and set to the equivalent of 20 keV. The ULD was set so as to avoid saturating the analog-to-digital (A/D) conversion. 

The Si(Li) electronics could be operated in two different trigger modes. The main mode was to accept a trigger from the TOF system to measure a charged particle track in coincidence. The Si(Li) could also be operated in self-trigger mode for X-ray calibrations and background measurements (ignoring the TOF trigger). In the TOF trigger mode, the card cages sent out a 15 $\mu$s long BUSY signal to the TOF electronics to inhibit further triggers until the data were processed and all A/D conversions had finished. The state FPGA also counted the total time and the time during which it was not busy to determine the livetime of the experiment.

At the end of each event,  the synchronized common 10 MHz clock counter (32~bit) was read out and merged into the event structure that contained the counter and ADC values. The DSP then sparsified the energy deposition data to reduce the overall data rate and sent out Ethernet packets using the UDP protocol to the flight computer. After each event the FPGAs were reset.
The DSP was also responsible for reading out housekeeping information from the card cages, e.g., voltage levels and temperatures and the register status of the state FPGAs.

\subsection{Flight Computer}
\label{fc}
The flight computer was mounted next to the TOF readout electronics in the pressure vessel and was composed of small form factor PC/104(+) modules. The computations were carried out on a low-power-consumption Pentium M 1.8 GHz CPU board with several fans for cooling. Communication to the Si(Li) tracker and the TOF readout systems was handled through the UDP Ethernet protocol via an extended-temperature-range Ethernet switch mounted outside of the pressure vessel.  The computer connected to the ISAS telemetry system via serial RS-232C interfaces. One telemetry link with 57.6 kbps was used to send data to ground and one link with 19.2 kbps was used to send commands to the payload. The RS-232C communication used an Xtreme/104+ Opto serial card (PC/104+, optically isolated serial interface), optically isolating the flight computer from the ISAS telemetry. A DIAMOND-MM-32DX-AT module was responsible for digital I/O operations like switching relays in the power systems, as well as for reading voltages, temperature sensors, and pressure transducers. In addition, voltages could be set to control the payload rotator and the cooling pump. 

The GPS antenna and fiber-optic gyroscope interfaced to the flight computer via two independent RS-232C connections. The ADCs of the monitoring board inside the power distribution box digitized the voltage and current levels and were read by a SUB-20 multi-interface USB adapter (DIMAX, SUB-20)  connected to the flight computer. The 10 MHz common clock was generated by a custom-made PC/104 board and sent to the tracker and TOF readout systems where the clock pulses were counted locally to generate event numbers. A much shorter synchronization pulse could be emitted from the same board to the individual readout systems to reset all different clock counters at the same time to 0. In this way events coming from the TOF and the two tracker DSPs could be merged with 100 ns precision, well below the expected maximum count rate during X-ray calibration runs (${\sim}$2 kHz). 

The operating system of the computer was Debian Linux, and the flight computer software followed a multi-threaded approach written in C++ utilizing the Boost thread library. Each interface (Ethernet, four times RS-232C, Sub-20, housekeeping information) thus had an individual thread running for it. 

The incoming data were stored in two different buffers (physics event data and housekeeping information) and data-handling threads passed these data along to different storage locations. Full copies of the data were stored on two different compact flash disks connected to the flight computer. In addition, these data were sent out over the Ethernet to the TOF were they were stored on an external USB flash drive for quick access after water recovery. 

During the flight, all housekeeping (${\sim}$1 kB/s) and about 10\% of flight data were sent over the telemetry link. Thus the health of the instrument was monitored at all times and as much science data as possible was sent for real-time analysis. The command execution was controlled by an individual thread and allowed the operator on the ground to control the experiment or adjust software parameters. All executed commands were mirrored back to ground by the flight computer in the regular data telemetry stream, together with an execution status. The attitude control system thread was part of the flight software and was set up as a feedback loop with the GPS antenna or gyroscope information. Furthermore, the CPU had hardware watchdog capabilities. A certain register had to be written with a frequency higher than 1 Hz or would otherwise trigger a reboot of the flight computer. It was also possible to send a special discrete command from the ground to reboot the computer during flight, but this option was not used during the actual flight. 

The ground computing environment was a graphical-user-interface application written in C++ using the QT and ROOT\cite{root} libraries. It allowed display of the raw physics event data and housekeeping information.

\section{Simulations}
Several distinct simulation efforts were used to optimize the design and performance of pGAPS. Those techniques were confirmed against the performance of the pGAPS instrument and will serve as the basis of bGAPS design and simulation efforts. 

\subsection{X-ray Tube Placement}
A GEANT4\cite{geant4} simulation was conducted to optimize the placement of the X-ray tube for detector calibration during the flight. All the X-ray interactions (scattering and photo-absorption) were taken into account in the simulation with the actual instrumental design including the Al housing, X-ray filter, and Si oil in the X-ray tube. This provided the X-ray spectrum and the count rate at each detector for the different types of filters available and mounting position options of the X-ray tube. The results were consistent with measurement on the ground and a location just under the bottom TOF paddles (and with an Al and Au filter)  was determined to be a favorable location for tube placement.

\subsection{Acceptance Simulation}
\label{rateSim}
\begin{figure}
  \centering
  \includegraphics[width=0.7\textwidth]{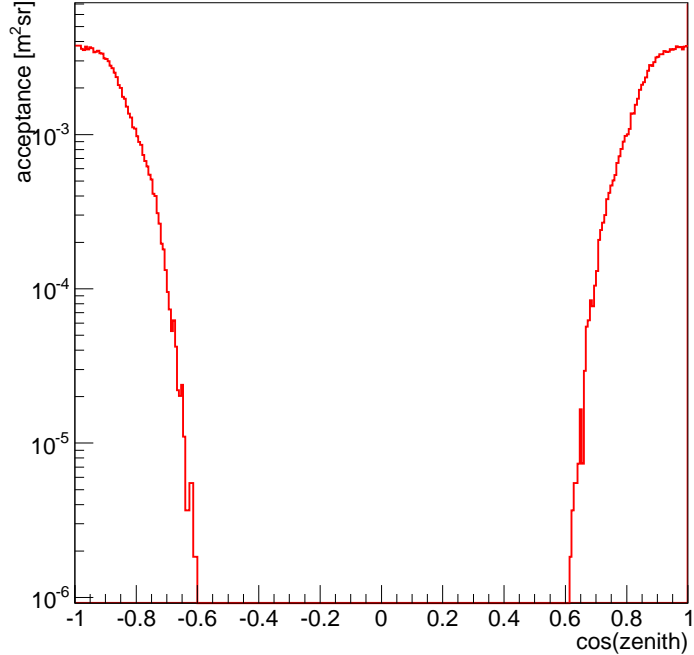}
	\caption{Monte Carlo estimate of the TOF trigger acceptance as a function of the cosine of the zenith angle.
}
\label{fig:pGAPS_tof_acceptance}
\end{figure}

The acceptance of pGAPS was calculated following the geometric approach described in \cite{sullivan}. Using the pGAPS geometry and the trigger condition explained in Sec. \ref{trs}, the total TOF trigger acceptance was 0.184 m$^2$sr and is shown as a function of the cosine of the zenith angle in Fig. \ref{fig:pGAPS_tof_acceptance}. As calculated with the PLANETOCOSMICS atmospheric simulation package\cite{pc}, particle tracks with any number of hits in the Si(Li) detectors were expected to accompany about 10\% of the TOF triggers.

\subsection{TOF plastic Scintillator Design}
\label{tofsim}
A dedicated simulation in GEANT4 was written to optimize the TOF plastic scintillator design. This simulation turned on the optical physics available in GEANT4 and the detector models had detailed optical properties attached to all components. This simulation was used to evaluate several light guide design changes, as well as to test several reflective wrapping material options. These simulations were cross-checked with prototype hardware in the lab to refine the model parameters. The model was validated, making it a valuable design tool for scintillator detector design work on bGAPS.

\section{The pGAPS Flight and Preliminary Analysis}
The pGAPS payload was successfully launched on June 3, 2012 at 4:55 a.m. JST from Taiki, Japan\cite{taiki}, which is on the east coast of the island of Hokkaido. The exact launch location was at a latitude of $42.50^{\circ}$ N and a longitude of $143.43^{\circ}$ E. The balloon drifted out over the Pacific at so-called “boomerang” altitudes of 12--15 km before releasing more ballast and going up to the “float” altitude of 33 km\cite{boomerang}. During that time of the year, the winds at these two altitudes blow in opposite directions, and from the moment the balloon reached float it was blown back towards the coast. The payload stayed higher than 30 km for 3 hours and 10 minutes. At 11:05 am JST the balloon was destroyed, and the gondola landed in the water at 11:35 am, where it was recovered within minutes by boat.

\subsection{Cooling System, OHP, and Detector Thermal Response}
Because pointing of the pGAPS payload was not available in fight (due to the failure of the rotator), the preflight thermal predictions required adjustment to correlate to the flight conditions.

\subsubsection{Radiator Thermal Response}
\begin{figure}
  \centering
  \includegraphics[width=0.7\textwidth]{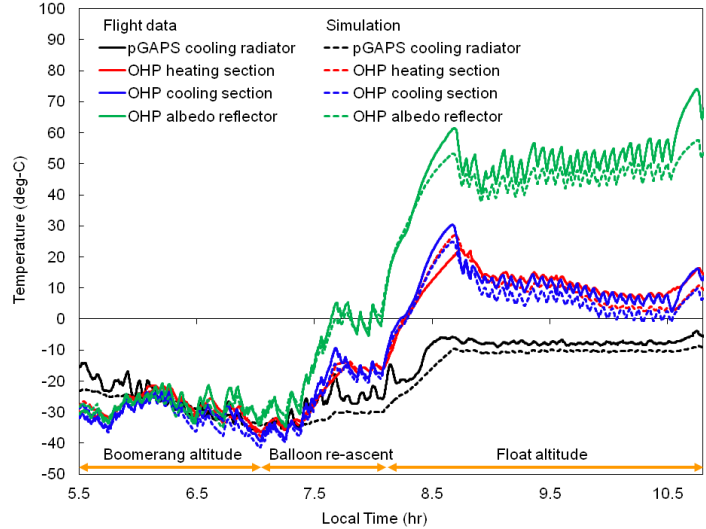}
	\caption{Measured temperatures from the pGAPS flight cooling radiator and thermal model output are shown (black lines). Also shown are temperatures of OHP heating section (red), cooling section (blue) and OHP albedo shield (support plate) (green). The thermal model reproduces the flight data very well.
}
\label{fig:thermalModel}
\end{figure}

A detailed parametric finite element model of the space radiator was created. In addition to the time-dependent ambient air temperatures, densities, and velocities; the random and near periodic solar gain and radiation cooling, caused by the rotation of the gondola, were included. To account for the loss of directional control during the flight, a detailed multiple load-step thermal-transient simulation was required. The period of rotation for the payload varied throughout the flight, with 10--15 minutes typical in the tropopause (boomerang altitude), $\sim5$ minutes at float altitude, and a period during assent to float where the radiator pointed at the sun. The 3 thermal regions were all treated with slightly different parameters to account for the different solar and albedo loadings and the change in convective thermal coupling. Solar gain parameters were: 0.5 for albedo, 250 K for ambient IR, and 1350 W/m$^2$ for the local solar intensity. Rotational orientation data were used to determine the time-dependent solar loading and cooling to space. The predicted (simulation) transient surface node temperature of the radiator and the actual surface temperature (flight data) of a mounted thermocouple are shown for the pGAPS flight in Fig. \ref{fig:thermalModel}. The thermal model for the payload radiator correlated well with the actual flight temperature data (lower black curves).

\begin{figure}
  \centering
  \includegraphics[width=0.7\textwidth]{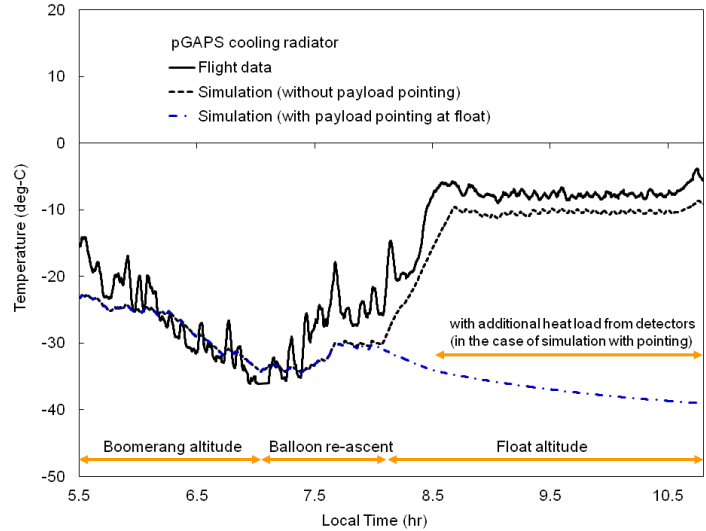}
	\caption{The predicted temperature of the pGAPS cooling radiator are shown if payload pointing was present. 
}
\label{fig:thermalWithPointing}
\end{figure}

The loading condition with the gondola (radiator) controlled to point in the antisolar direction at float altitude was simulated (Fig. \ref{fig:thermalWithPointing}). The actual expected dissipated 1.2 W thermal load of the electronics from the six active detectors was included at float. At float, there is a time lag of about an hour before the temperature of the detectors will reach the required -35 C. This is caused by the heating of the radiator during ascent from the tropopause (no directional control) with the radiator side of the gondola (worst-case condition) primarily facing the sun.

\subsubsection{Oscillating Heat Pipe Test Unit Performance}

Fig. \ref{fig:thermalModel} show the temperature profile of the prototype OHP at boomerang and float altitudes. The dashed lines in Fig. \ref{fig:thermalModel} show temperatures calculated by solving the transient heat equation using a three-dimensional finite-element method. In the simulation, almost all thermal characteristic variables were either measured in our laboratory or taken from the literature. Only the surface characteristics of the albedo reflector (an aluminum plate), which might be partially covered by frost during the flight, and the heat-transfer coefficient of the rarefied atmosphere, which is poorly understood, were modified to fit the simulation results to the flight data. The flight data is well reproduced by the simulation. The heat conductance of OHP was around 5 W/K at both altitudes, which indicates that the prototype OHP worked well during the flight as expected.

\subsubsection{Detector Thermal Response}
The original thermal model had boundary conditions defined by the time-dependent inner surface temperature of the coolant heat exchanger coupling for each Si(Li) detector layer. During the actual flight, the pump was not used due to the radiator being thermally uncontrolled (caused by the rotation of the gondola). This condition was inconsistent with the original SINDA\cite{sinda} thermal model, so the model was revised to represent the actual flight conditions. The detector vessel transient thermal response was simulated with a lumped parameter model. Radiative coupling between the gondola foam insulation and the detector vessel was included. Inside the detector vessel, the model incorporated the representative thermal masses for the detectors and major structural components. Additionally, the thermal mass of the static Fluorinert fluid and support tubing were present. The predicted transient temperatures of the silicon wafer detectors were then correlated with the pGAPS flight data.

\begin{figure}
  \centering
  \includegraphics[width=0.7\textwidth]{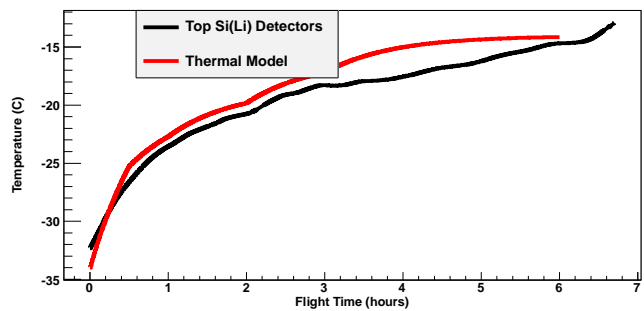}
	\caption{The measured and predicted temperature for a pGAPS Si(Li) top layer detector is shown.
}
\label{fig:tempTopDet}
\end{figure}

The predicted temperature of the thermal model with the modified ‘flight’ boundary conditions correlated well with the actual flight data. The temperature curves for detectors track the temperature data from the flight (See Fig. \ref{fig:tempTopDet}). Sensitivity studies indicate the importance of the detector vessel insulation and the dominance of radiation exchange between the inner wall of the gondola insulation and the surface of the insulated vessel.

\subsection{Si(Li) detector performance}

\begin{figure}
\centering
\begin{minipage}{0.5\textwidth}
  \centering
  \includegraphics[width=0.9\linewidth]{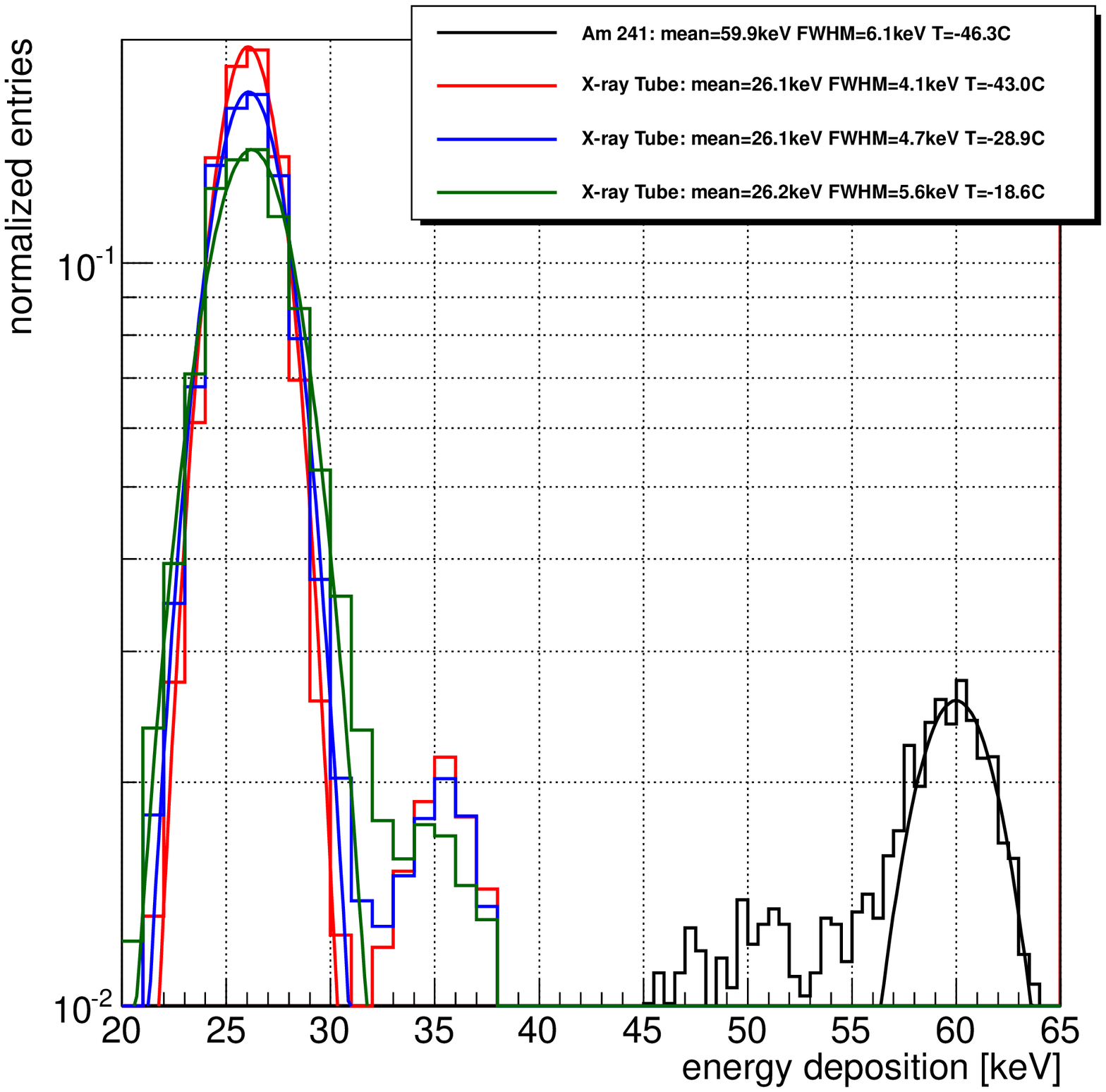}
\end{minipage}%
\begin{minipage}{0.5\textwidth}
  \centering
  \includegraphics[width=0.9\linewidth]{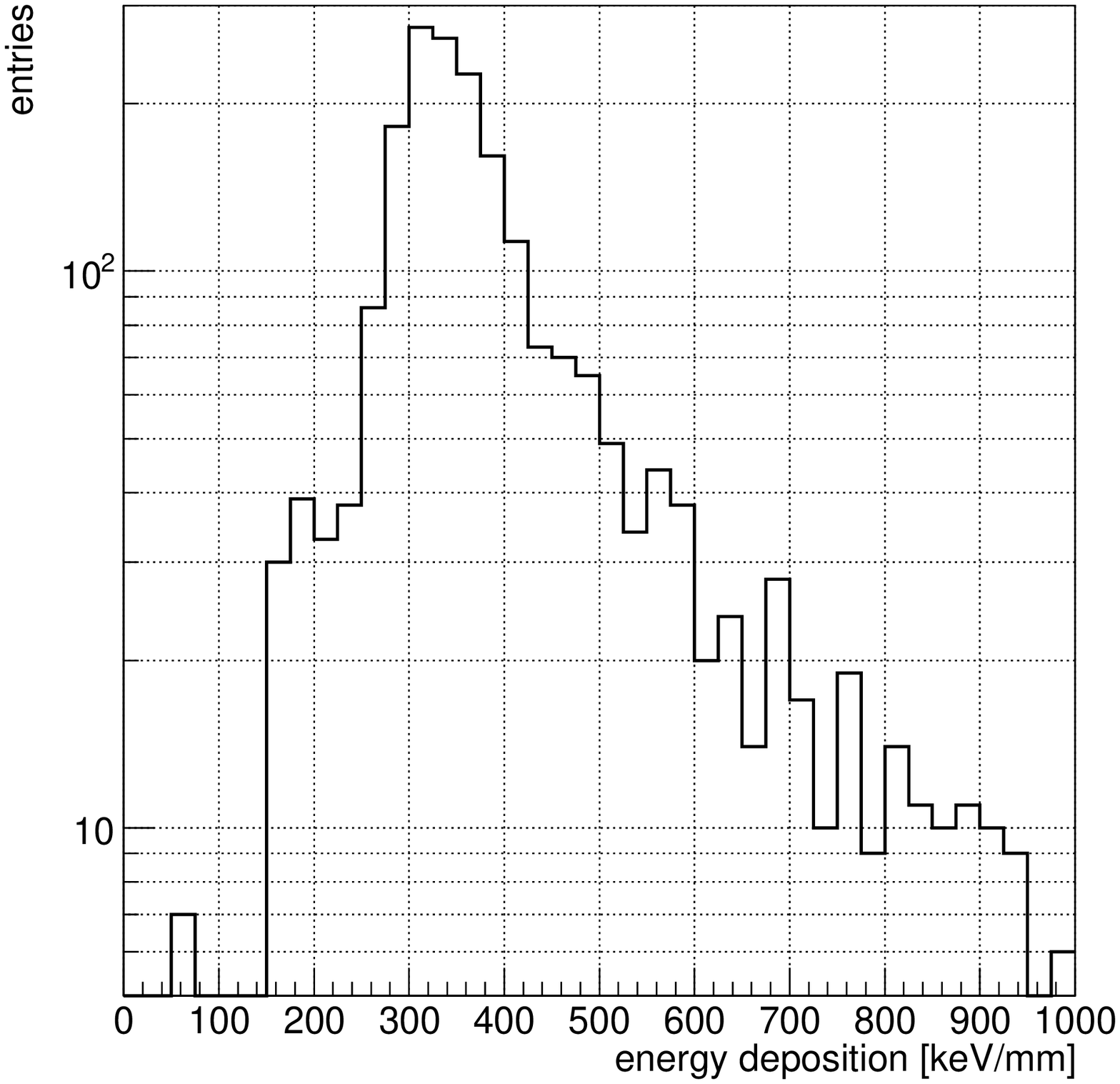}
\end{minipage}
\caption{Left: X-ray tube (preparation/red, right after launch/blue, float/green) and Am-241 source (preparation/black) measurements for one detector summed over all eight channels. Right: Charged particle energy deposit/mm of clean tracks in the Si(Li) detectors at float.
 }
\label{fig:siliFigs}
\end{figure}

During the launch preparations the tracker system was cooled down to an average temperature of -46 $^{\circ}$C inside the vessel using the liquid nitrogen system. The flight readiness procedure required that the  gondola be detached from any ground support equipment, including the cooling system, for about 25 minutes before the actual launch. During that time the temperature inside the vessel warmed up to about -35 $^{\circ}$C, which is the optimal operating temperature for the Si(Li) detectors. 

During the ascent phase, the gondola was expected to rotate freely, since the rotator was not designed to be strong enough to point the instrument in the denser air at lower altitudes. Due to the rotator failure, the tracker system continued to warm up at float. When the float altitude was reached, the mean temperature in the vessel was about -19 $^{\circ}$C and by the end of the flight it had increased to around -15 $^{\circ}$C. It was known from laboratory tests that some of the detector strips depended more strongly on the temperature than average, because of differences in the silicon surfaces and the grooves between strips. However, even at the very high temperature reached by the end of the flight, more than 60\% of the channels were working reliably. Several strips also had known bad bonding.

Measurements with the X-ray tube and an Am-241 source were carried out (Fig. \ref{fig:siliFigs}, left) and were used to calibrate the energy depositions for each detector strip. The dominant peak at  26 keV (with some contamination from a line at 22 keV) can be used as an indicator for the measurement stability and shows no shift of the peak position. A small widening of the distribution is visible and is in agreement with measurements on ground (due to the increasing operating temperature). The right side of Fig. \ref{fig:siliFigs} shows the distribution of charged particle energy deposition for hits on clean particle tracks which show a Landau distribution, as expected.

A sample of three of the flown SEMIKON Si(Li) detectors were operated after flight and their performance showed no degradation. This demonstrates that the these type detectors are quite robust even when not passivated for environmental protection.

\subsection{TOF detector performance}
\begin{figure}
\centering
\begin{minipage}{0.49\textwidth}
  \includegraphics[width=1.0\linewidth]{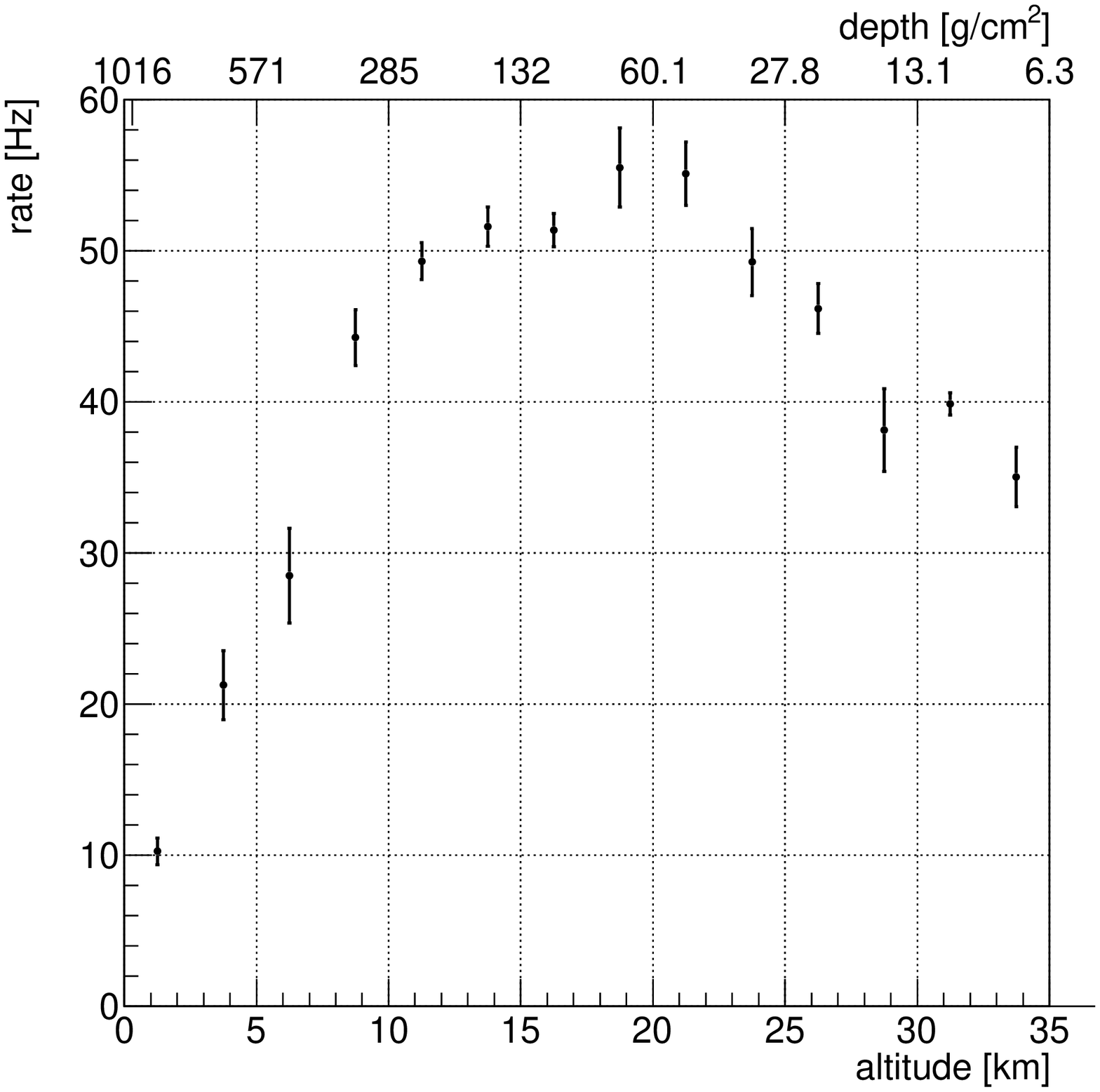}
  \caption{The pGAPS trigger rate recorded in flight shown as a function of altitude (and atmospheric depth).}
  \label{fig:tofTrigRate}
\end{minipage}\hfill
\begin{minipage}{0.49\textwidth}
  \includegraphics[width=1.0\linewidth]{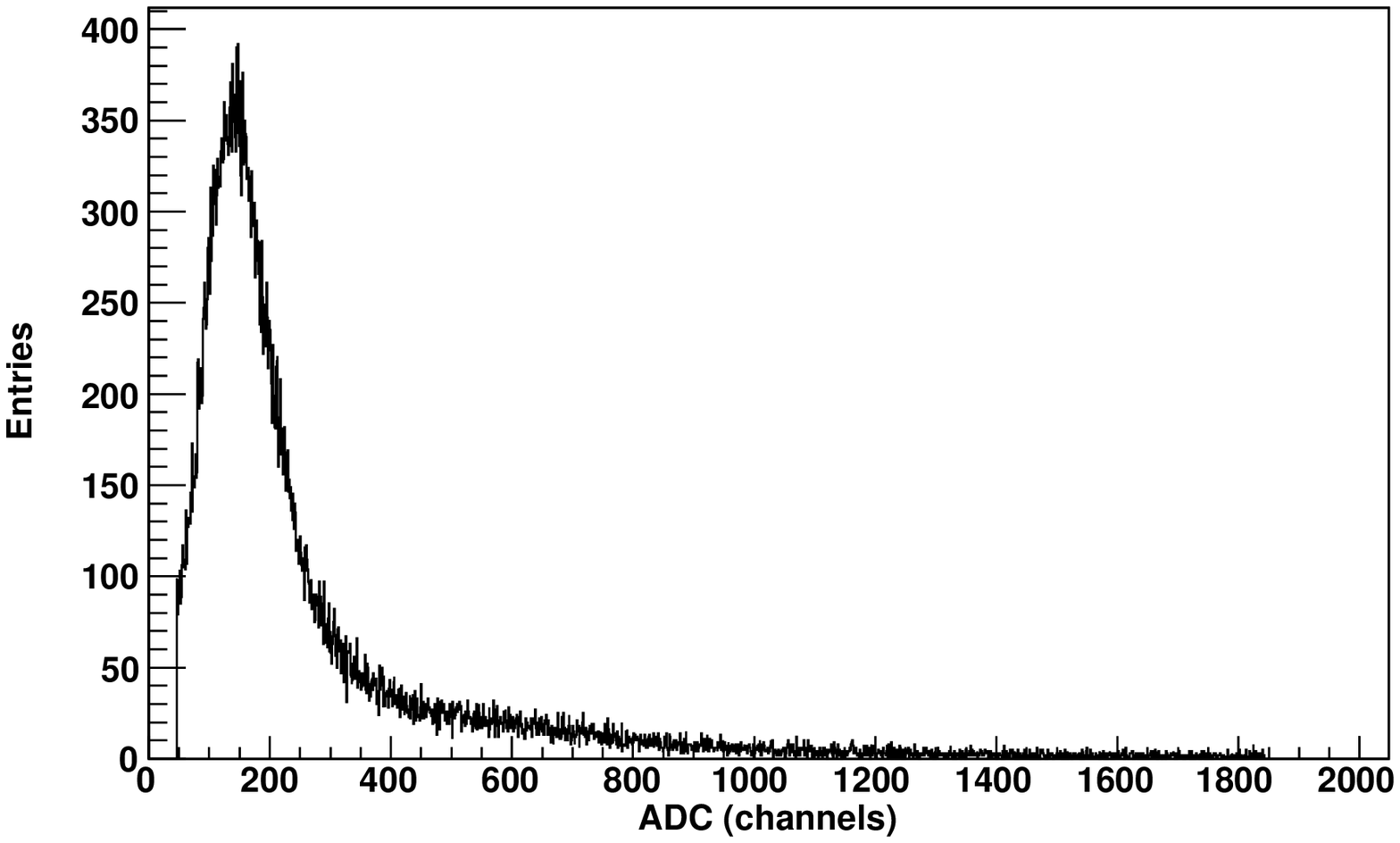}
  \caption{Uncorrected ADC response for tube 1 in the pGAPS TOF at float. The large peak centered near channel 200 (20 pC of ADC integration) corresponds to singly charged particles. This distribution is a typical distribution for all tubes, with some variation in gain between channels.}
  \label{fig:tofADC}
\end{minipage}
\end{figure}

The TOF system was operated continuously (except for a few system resets) from before payload launch until just before termination. All but one PMT (out of 32) operated properly for the duration of the flight, but one (tube 28 in the middle layer) did show signs of intermittent corona discharge upon reaching float altitude. Since each paddle is instrumented with two PMTs, the overall instrument acceptance and trigger efficiency was not reduced. The TOF provided the trigger to the payload (except when the Si(Li) readout was operated in self-trigger mode during X-ray tube calibration runs). The trigger rate as a function of altitude is shown in Fig. \ref{fig:tofTrigRate}. This rate agrees (to within a factor of 2) with the predicted rates discussed above in section \ref{rateSim}. Improved understanding of the instrument readout livetime is expected to better harmonize these numbers. 

The temperatures of the exposed TOF components stayed within acceptable ranges throughout the flight. At boomerang altitude, the temperatures measured on the PMTs and the preamplifiers dropped to between -15 $^{\circ}$C and -20 $^{\circ}$C, but stabilized back to ${\sim}0~^{\circ}$C at flight termination.

As shown in Fig. \ref{fig:tofADC}, the ADC measurements from the PMT anode taps showed singly charged particles, whose distribution peaked at a convenient part of the overall dynamic range. Hints of helium events are also visible. This distribution has no corrections for path-length attenuation or track-angle. Once those corrections have been made, the charge resolution is expected to improve.

\begin{figure}
\centering
\begin{minipage}{0.49\textwidth}
  \centering
  \includegraphics[width=0.9\linewidth]{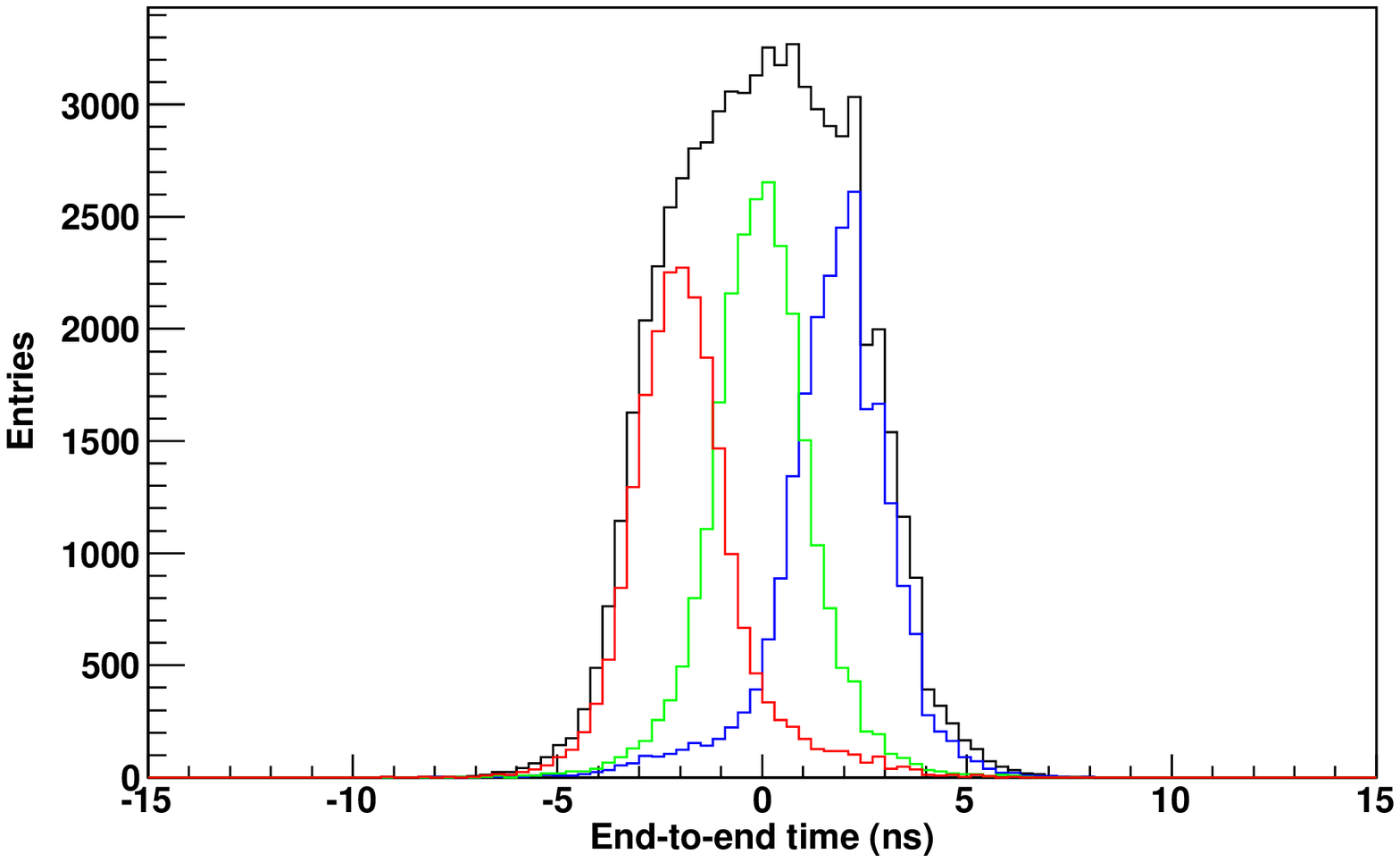}
  \caption{TDC time difference for paddle 1 in the pGAPS TOF. The time difference is proportional to the hit position down the length of the paddle. The three colored distributions shown are hits requiring a hit in a crossed paddle below in the same layer (different color for each of the crossed paddles).}
  \label{fig:tofEnd2end}
\end{minipage}\hfill
\begin{minipage}{0.49\textwidth}
  \centering
  \includegraphics[width=0.9\linewidth]{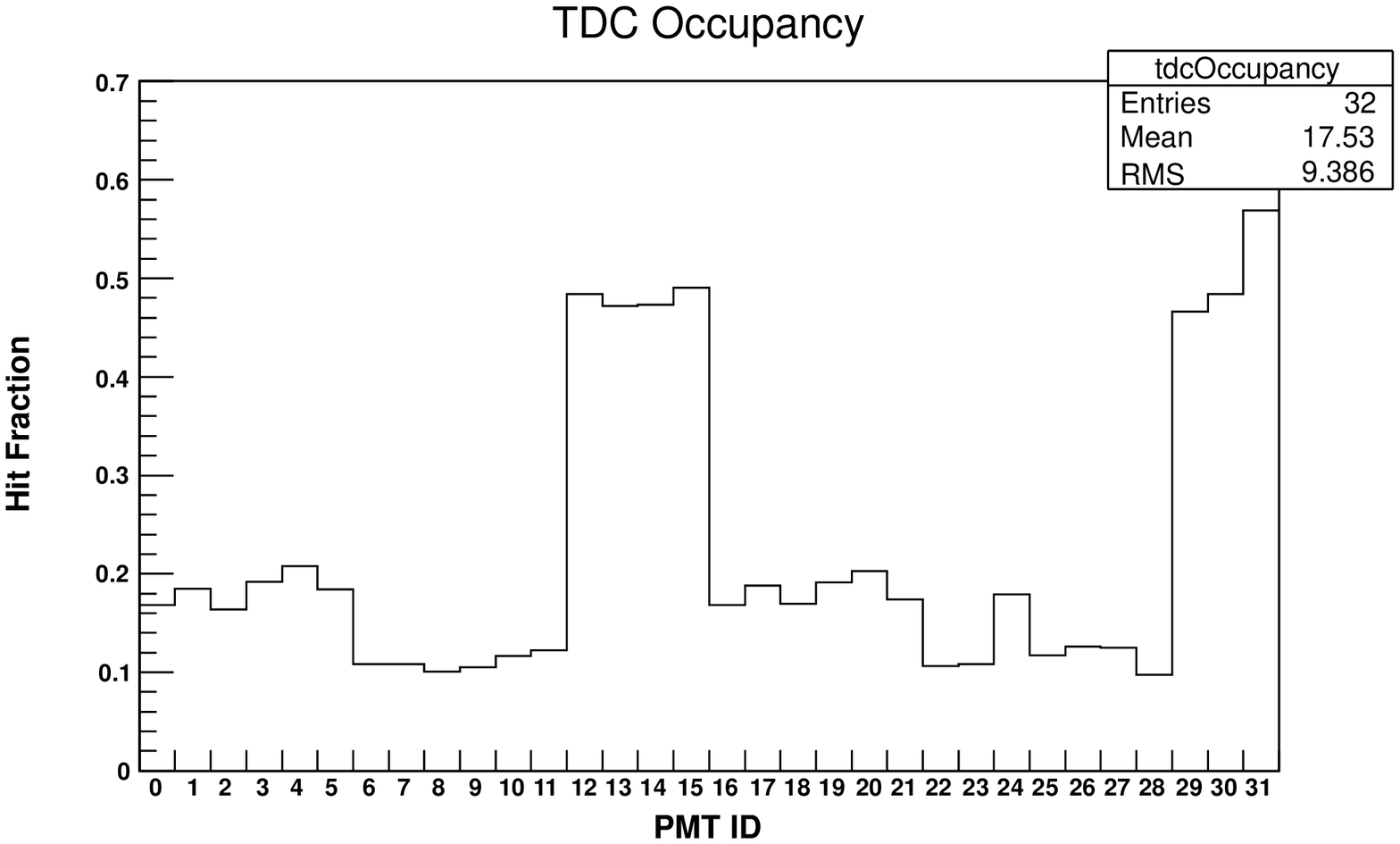}
  \caption{Hit fraction for each PMT in the pGAPS TOF for flight data. Tubes 0--5 and 16--21 were in the top layer, 12--15 and 28--31 were in the middle layer, 6--11 and 22--27 were in the bottom layer.}
  \label{fig:tofOccupancy}
\end{minipage}
\end{figure}

The TDCs reading out the PMT dynode taps also behaved as expected during the flight. Detailed work is continuing on finding all channel-to-channel offsets for the TDCs as well as slewing corrections to correct for the fixed-threshold response of the discriminators. Fig. \ref{fig:tofEnd2end} shows the difference of the two raw TDC signals attached to paddle 1. The TDC times have been converted from channels to ns, but no other corrections have been applied. This TDC difference is proportional to the hit position.

To compare the gain matching of all PMTs, Fig. \ref{fig:tofOccupancy} shows the TDC occupancy for all flight events. The trigger conditions used in flight required one x-going and one y-going PMT in the middle layer to fire (together with at least one more PMT from either the top or the bottom) which explains why the occupancy for middle-layer PMTs are all around 0.5 (except for tube 28, which showed signs of HV breakdown at float altitude). 

\section{Next steps for GAPS}
Following the successful flight of the pGAPS payload, the next logical step is the construction of the bGAPS experiment to search for cosmic ray antideuterons. Work is progressing on a design for this instrument, with a series of Antarctic science flights proposed for later this decade.

This instrument would have a 2 m $\times$ 2 m $\times$ 2 m central cube of Si(Li) detectors ($\sim13$ layers, several thousand detectors), all surrounded by a large area time-of-flight detector. An instrument of this physical size and channel count presents a number of engineering challenges, but the performance of pGAPS shows the design approach is sound and can be scaled to a larger instrument.

The basic designs for the TOF scintillators would only require some scaling to a larger size (from 0.5 m to 1 m). The Si(Li) detectors would not require significant modification, however, due to the large number needed, they must be produced in-house for the experiment to be economical. The production process is well understood however and is currently being implemented. Development of a highly integrated readout system for both detector systems will be needed however, with optimizations for size, mass, performance, and power consumption. Modern application-specific integrated circuits offer many (commercially available) options for very high-performance readout schemes with excellent performance, high channel count, and low power consumption.

As discussed above, the measurements from bGAPS would not only probe areas of parameter space complementary to other (both direct and indirect) searches, but also could be invaluable for confirming a detection by other experiments.

\section{Conclusions}
The pGAPS experiment had an extremely successful test flight in 2012, and all major mission goals were met. All science detectors performed as expected, and over 1 million cosmic ray triggers were recorded. Additionally, a large number of X-ray calibrations runs were performed during the flight to characterize Si(Li) detector performance.

Data was also collected to flight qualify our cooling approach and validate and improve our thermal model, which is essential for an instrument with detector cooling and thermal regulation required. Additionally, a very promising alternative cooling technique (the OHP) was tested for the first time in a balloon flight environment. 

Analysis of the flight data is still progressing, but early results indicate the characterizations of the X-ray and charged cosmic ray backgrounds have been fully successful. Additionally, performance of all detector components was very good, and provides a clear path for development of the bGAPS payload instrument. A much more detailed discussion of performance and science results from the pGAPS flight will be presented in a separate analysis paper which is being prepared by the collaboration and will be published shortly.

\section*{Acknowledgements}
Funding for the pGAPS project has been provided by NASA grants in the US, by MEXT-KAKENHI grants in Japan, and the UCLA Division of Physical Sciences. This material is based upon work supported by the National Science Foundation under Award No. 1202958. Additionally, we would like to thank Dennis Stefanik, Paul Kaplan, Connor Hailey, Jane Hoberman, Brent Mochizuki, Katayun Kamdin, Mayra Lopez-Thibodeaux, and Tracy Zhang for their contributions to the project.





\bibliographystyle{elsarticle-num}
\bibliography{pGAPSinstrumentMognet}







\end{document}